\documentclass[twocolumn,superscriptaddress,showpacs,amsfonts,nopacs,amsmath,amssymb]{revtex4-1}
\pdfoutput=1

\usepackage{epsfig}
\usepackage{graphicx}
\usepackage{longtable}
\usepackage{color}

\usepackage{bm}

\usepackage{amsmath,amssymb,mathrsfs}
\usepackage{mathtools}
\usepackage{braket}
\usepackage{setspace}
\usepackage[table]{xcolor}

\newcommand{\bea}{\begin{eqnarray}}
\newcommand{\eea}{\end{eqnarray}}

\begin{document}

\title{Tuning nodal line semimetals in trilayered systems}
\author{Filomena Forte}
\affiliation{Dipartimento di Fisica ``E.R. Caianiello'',\\
Universit\`a degli Studi di Salerno, Via Giovanni Paolo II, 132, 84084 Fisciano (SA), Italy}
\affiliation{CNR-SPIN, UOS di Salerno, Via Giovanni Paolo II, 132, 84084 Fisciano (SA), Italy}
\author{Delia Guerra}
\affiliation{Dipartimento di Fisica ``E.R. Caianiello'',\\
Universit\`a degli Studi di Salerno, Via Giovanni Paolo II, 132, 84084 Fisciano (SA), Italy}
\author{Canio Noce}
\affiliation{Dipartimento di Fisica ``E.R. Caianiello'',\\
Universit\`a degli Studi di Salerno, Via Giovanni Paolo II, 132, 84084 Fisciano (SA), Italy}
\affiliation{CNR-SPIN, UOS di Salerno, Via Giovanni Paolo II, 132, 84084 Fisciano (SA), Italy}
\author{Wojciech Brzezicki}
%
\affiliation{International Research Centre MagTop at Institute of Physics, Polish Academy of Sciences, Aleja Lotników 32/46, PL-02668 Warsaw, Poland}
\author{Mario Cuoco}
\affiliation{CNR-SPIN, UOS di Salerno, Via Giovanni Paolo II, 132, 84084 Fisciano (SA), Italy}
\affiliation{Dipartimento di Fisica ``E.R. Caianiello'',\\
Universit\`a degli Studi di Salerno, Via Giovanni Paolo II, 132, 84084 Fisciano (SA), Italy}
\begin{abstract}
We investigate two-dimensional trilayered quantum systems with multi-orbital conduction bands by focusing on the role played by the layer degree of freedom in setting the character of nodal line semimetals. The layer index can label the electronic states where the electrons reside in the unit cell and can enforce symmetry constraints in the electronic structure by protecting bands crossing.
We demonstrate that both the atomic spin-orbit coupling and the removal of local orbital degeneracy can lead to different types of electronic transitions with nodal lines that undergo a changeover from a loop structure enclosing the center of the Brillouin zone to pockets winding around multiple high symmetry points. We introduce and employ a criterion to find the nodal lines transitions. On the basis of a zero-dimensional topological invariant that, for a selected electronic and energy manifold, counts the number of bands below the Fermi level with a given layer inversion eigenvalue in high symmetry points of the Brillouin zone, one can determine the structure of the nodal lines and the ensuing topological transitions.    
\end{abstract}

\maketitle

{\section{Introduction}}

Recently, the theoretical prediction \cite{Kane2005,Bernevig2006,Moore2007,Fu2007}
and experimental achievement \cite{Konig2007,Hsieh2008,Xia2009a}
of topological insulators due to strong spin-orbit coupling
(SOC) have dramatically enriched the scenario of phases of matter which can be obtained by suitably designing quantum materials. 
Apart from topological insulators~\cite{Qi2011,Hasan2010}, there has been a significant expansion towards topologically protected gapless phases, e.g. metals and semimetals \cite{Volovik2003,Volovik2007,Wan2011,Heikkila2011,Yang2011,Burkov2011,Krempa2012,Xu2011,Halasz2012}, thus
boosting the discovery of novel materials~\cite{Weng2015b,Huang2015,Xu2015a,Lv2015a,Lv2015b,Xu2015b}
with non-trivial band crossing points in the momentum space as well as quantum materials that combine topological and conventional forms of ordering. 
On a general ground, topological semimetals~\cite{Burkov2016} are materials where conduction and valence bands exhibit crossings in some points or lines in the Brillouin zone and the crossings can occur as protected by certain symmetry of the system or by the presence of topological invariants. Among the available topological gapless states, the Dirac semimetals owe a particular interest, with massless Dirac fermions emerging as low-energy fundamental excitations. 
Since the Dirac semimetals have an intrinsic instability, a symmetry protection is needed ~\cite{Yang2014,CastroNeto,Young2012}, as, for instance, it occurs in graphene, and, more importantly, it is also the dimensionality of the electronic environment that plays a crucial role in setting the robustness and the character of the semimetallic phase.  

When considering low-dimensional materials, the physical properties can be critically linked to the number of layers in the unit cell and, hence, to the given effective dimensionality of the electronic environment. Another important aspect of quasi two-dimensional quantum materials is represented by the emergence of a layer degree of freedom which can be also crucial in determining the character of the electronic states where the electrons are located in the unit cell. For instance, such an internal degree of freedom, in analogy with electron spin or the orbital pesudospin, can be employed as a carrier of classical or quantum information and thus it may be a relevant ingredient to explore the design of electronic device directions. Relevant proposals along this direction are represented, among the many, by the graphene bilayers \cite{San-jose2009,Pesin2012} and metal dichalcogenide bilayers \cite{Gong2013}.
In this context, trilayered systems can be a quite unique system because they are marked by inequivalent layer components when considering the possible mismatch of the inner and outer layers in the unit cell. Concerning the materials perspective, such three-layers design can be easily accessed in the realm of oxide materials both by confining few unit cells between insulating materials (i.e. SrTiO$_3$ or LaTiO$_3$ ) or, for the case of quasi 2D systems, by considering the members of the Ruddlesden-Popper (RP) series A$_{n+1}$B$_n$O$_{3n+1}$, with A being the alcaline or rare earth element and B the transition metal atom.   

In this paper, we investigate the nature of the nodal line states occurring in trilayered quantum systems with multi-orbital configurations at the Fermi level and spin-orbit coupling.
We employ the symmetry associated with the inversion of the layer index to label the electronic states and single out the ensuing protected bands crossing.
The investigation of the electronic structure indicates that both the atomic spin-orbit coupling and the local removal of orbital degeneracy can lead to different types of electronic topological transitions with nodal lines bands crossing that get converted from a loop structure enclosing the center of the Brillouin zone to pockets winding around multiple high symmetry points. We find a general criterion to find the nodal lines semimetal transitions in the parameters space. Indeed, on the basis of a zero-dimensional topological invariant that, for a selected electronic and energy manifold, counts the number of bands below the Fermi level with a given layer mirror eigenvalue in high symmetry points of the Brillouin zone, one can determine the overall structure of the nodal lines and the ensuing topological transitions.    

The paper is organized as follows. In Sec. II we present
the model Hamiltonian employed to describe the trilayered system and the ensuing symmetry properties. Sec. II is devoted to the phase diagram and the analysis of the nodal line semimetal emerging among the obtained electronic
structures. In Sec. IV we provide the summary and the concluding remarks.

{\section{Model Hamiltonian, symmetry relations and local energy gaps}}

\subsection{Model Hamiltonian and symmetries}

The model Hamiltonian for the 2D trilayered system with tetragonal symmetry is based on three atomic orbitals (e.g. t$_{2g}$ or $p$ bands) describing itinerant electrons in the
presence of atomic spin-orbit coupling and an effective layer-dependent crystal field potential that removes the local orbital degeneracy. Hereafter, we assume that we deal with a t$_{2g}$ quantum material.
Then, the Hamiltonian can be written as:
\begin{equation}
\label{HamiltTr}
H_k=H_{t_p}+H_{t_o}+H_{\Delta_I}+H_{\Delta_O}+H_{SOC} \,.
\end{equation}
The first term in \eqref{HamiltTr} represents the electronic hopping in the $xy$ plane, assuming that it is limited to nearest-neighbors only, it is equal to:
\begin{eqnarray}
H_{t_p}=&&\sum_\bold{k} [-4 t_{xy}(\cos{k_x}+\cos{k_y})-4t_{xz}\cos{k_x}+ \nonumber
\\ && -
4t_{yz}\cos{k_y}] c^\dagger_\bold{k}c_\bold{k}
\end{eqnarray}

The next term of \eqref{HamiltTr} is constituted by the intra-cell hopping, which only involves charge transfers between the $\gamma$z orbitals $(xz,yz)$ within different layers. It can be written as follows:
\begin{equation}
\label{orthop}
H_{t_o}=t_o \sum_{\alpha=\gamma z ,  \sigma} c^\dagger_{\alpha  \sigma} c_{\alpha  \sigma} + {\text{h.c.}} \,.
\end{equation}

The other contributions are related to the on-site energies, which simulate the crystal field splitting in the trilayered structure and can be expressed as:
\begin{equation}
\label{cftri}
H_{\Delta_I}= \sum_{\alpha,\sigma} \varepsilon_\alpha n_{\alpha_1 \sigma}  \qquad H_{\Delta_O}=\sum_{\alpha,\sigma} \varepsilon_\alpha n_{\alpha_2 \sigma} +\sum_{\alpha,\sigma} \varepsilon_\alpha n_{\alpha_3 \sigma}  
\end{equation}
where $\alpha$ labels the $xy$ and $\gamma$z orbitals, $\sigma$ is the spin index and the indices 1,2,3 are related to the inner and the outer layers of the three-layered structure, respectively. 
 
The SOC Hamiltonian can be expressed as:
\begin{equation}
\label{SOCtri}
H_{SOC}=\lambda  \bold{l}\cdot \bold{s} \,.
\end{equation}
We denote by the 3-vector of $2\times2$ matrices $\bf{s}=\frac{1}{2}\bf{%
\sigma }$ is the spin operator expressed through the Pauli matrices $%
\bf{\sigma }$. 
Furthermore  the 3-vector of $3\times 3$ matrices $\bf{l}$ is the projection of the orbital angular momentum
operator to the t$_{2g}$ subspace. It has components 
$(l_{k})_{\alpha \beta }=\mathrm{i}\epsilon_{k\alpha \beta }$, $k=x,y,z$,
such that 
$\bf{l}\times \bf{l}=-\mathrm{i}\,\bf{l}$. 
Explicitly, in the basis $(d_{yz},d_{xz},d_{xy})$, the matrices for the orbital operators are
\begin{equation}
l_{x} =%
\begin{pmatrix}
0 & 0 & 0 \\ 
0 & 0 & \mathrm{i} \\ 
0 & -\mathrm{i} & 0%
\end{pmatrix},%
\ l_{y} =%
\begin{pmatrix}
0 & 0 & -\mathrm{i} \\ 
0 & 0 & 0 \\ 
\mathrm{i} & 0 & 0%
\end{pmatrix},%
l_{z} =%
\begin{pmatrix}
0 & -\mathrm{i} & 0 \\ 
\mathrm{i} & 0 & 0 \\ 
0 & 0 & 0%
\end{pmatrix}.
\end{equation}

Since we have three layers and nine local atomic basis, the eighteen configurations can generate the overall basis set: [$xy_{1\uparrow}, xz_{1\uparrow}, yz_{1\uparrow}, xy_{2\uparrow},xz_{2\uparrow},yz_{2\uparrow},xy_{3\uparrow}$,$xz_{3\uparrow},yz_{3\uparrow},xy_{1\downarrow}, xz_{1\downarrow}$, $yz_{1\downarrow}, xy_{2\downarrow},
xz_{2\downarrow},yz_{2\downarrow},xy_{3\downarrow},xz_{3\downarrow},yz_{3\downarrow}$], 
where the index 1 stands for the inner layer, the indices 2 and 3 are related to the outer layers. 
The in-plane hopping matrix can be written in a block matrix structure:

\begin{equation}
H_{t_p}=
 \begin{pmatrix}
P& 0 &0&0&\dots& 0\\
0&P&0&0&\dots&0\\
0&0&P&0&\dots&0\\
0&\dots&0&P&0&0\\
0&\dots&0&0&P&0\\
0&\dots&0&0&0&P
\end{pmatrix}
\end{equation}
where $P$ is a 3$\times$3 matrix equal to:
\begin{equation}
P=
\begin{pmatrix}
-4t_{xy} (\cos k_x + \cos k_y)&0&0\\
0&-4t_{xz} \cos k_x&0\\
0&0&-4t_{yz} \cos k_y \,.
\end{pmatrix}
\end{equation}
Hereafter, we assume t$_{xy}$=t$_{xz}$=t$_{yz}$=t$_p$.\\
The intra-cell hopping matrix has the following structure:
\begin{equation}
H^\sigma_{t_o}=
\begin{pmatrix}
0&0&0&0&0&0&0&0&0\\
0&0&0&0&t_o&0&0&t_o&0\\
0&0&0&0&0&t_o&0&0&t_o\\
0&0&0&0&0&0&0&0&0\\
0&t_o&0&0&0&0&0&0&0\\
0&0&t_o&0&0&0&0&0&0\\
0&0&0&0&0&0&0&0&0\\
0&t_o&0&0&0&0&0&0&0\\
0&0&t_o&0&0&0&0&0&0\\
\end{pmatrix}
\end{equation}
where H$_{t_o ij} \neq$ 0 if i and j correspond to two analogous $\gamma$z orbitals of different layers.

The CF contribution\eqref{cftri} assumes the form:

\begin{equation}
\label{hamD}
H_{\Delta_I}+H_{\Delta_O}= \begin{pmatrix}
D1& 0 &0&0&\dots& 0\\
0&D2&0&0&\dots&0\\
0&0&D3&0&\dots&0\\
0&\dots&0&D1&0&0\\
0&\dots&0&0&D2&0\\
0&\dots&0&0&0&D3
\end{pmatrix}
\end{equation}
with
\begin{eqnarray}
D1=&&
\begin{pmatrix}
\Delta_I&0&0\\
0&0&0\\
0&0&0
\end{pmatrix} \quad
D2=
\begin{pmatrix}
\Delta_{O2}&0&0\\
0&0&0\\
0&0&0
\end{pmatrix}  \nonumber
\\ 
D3=&&
\begin{pmatrix}
\Delta_{O3}&0&0\\
0&0&0\\
0&0&0
\end{pmatrix}
\end{eqnarray}
In the above expressions, $\Delta_{O2}$ and $\Delta_{O3}$ are the crystal field potentials that split the $xy$ and $\gamma$z orbitals of the outer layers, while $\Delta_I$ is the corresponding quantity relative to the inner layer. In tetragonal perovskite oxides, it the compression/elongation of the octahedra along the $z$ axis that can lead to inequivalent removal of the orbital degeneracy in the outer and inner layers. Moreover, in writing \eqref{hamD}, we assume that the energy of the t$_{2g}$ orbitals is measured with respect to the $\gamma$z orbitals, then $\Delta_I$ is identified with the energy of the $xy$ orbital of the inner layer, while $\Delta_{O2}$ and $\Delta_{O3}$ represent the $xy$ orbital energies of the outer layers.\\ 

Concerning the symmetry properties of the model Hamiltonian, one can show that the usual time reversal operator, $T=i \sigma_y K$, generates the time-symmetry transformation, with $K$ being the complex conjugation operator. Moreover, the system is inversion symmetric via the unitary operator $\mathscr{P}$ that invert the sign of the in-plane spatial coordinates. Then, the $\mathscr{P} T$ combination allows to have an antiunitary symmetry that is local in momentum space and implies that all the eigenstates are twofold degenerate in the Brillouin zone.

Apart from the time and inversion symmetry, if the crystal field potentials at the outer layers have the same amplitude, i.e. $\Delta_{O2}=\Delta_{O3}$, the Hamiltonian owes an additional symmetry $\mathcal{F}$ associated to the inversion of the outer layers within the unit cell. Since $\mathscr{F}^2$=1, the eigenstates of the Hamiltonian can be labelled by its $\pm$ eigenvalues and construct two subspaces whose wavefunctions have opposite parity:
\begin{equation}
\label{mirrortr}
\mathscr{F} \ket{\Psi}=-\ket{\Psi}, \qquad \mathscr{F} \ket{\Psi}=\ket{\Psi} \,.
\end{equation}

The states for which $\mathscr{F}$=-1 are odd and have a non-bonding character, while the states with $\mathscr{F}$=1 are even and have bonding or anti-bonding character.
In summary, by diagonalizing the Hamiltonian \eqref{HamiltTr} we obtain nine energy bands which are doubly degenerate as due to the time-inversion symmetry. If the outer layers are equivalent, according to the layer-interchange symmetry, one can further classify the nine bands by means of the layer parity index $\textit{f}$, associating to each of them $\textit{f}=\pm$1 depending on the eigenvalue of the $\mathscr{F}$ symmetry. 
\\
\subsection{Local energy gaps: interplay of intra-unit cell hopping, spin-orbit and crystal field potential}

In this subsection, we investigate the energy relations of the states within the unit cell by assuming the in-plane hopping amplitude is vanishing, $t_p=0$. This regime of low electronic connectivity is relevant when the intra-unit cell energies are dominant due to geometrically induced weak orbital overlaps in the plane as well as electron-electron correlation effects. For instance, in the case of 2D tetragonal perovskite, small amplitude of the inter-unit cells charge transfer can arise in the presence of large rotations of the octahedra, i.e. strongly distorted bonds, or by considering a larger separation between the atoms whose orbitals are mainly contributing to the conduction bands, as for the case of double perovskite systems.

\begin{figure}[h]
\noindent \begin{centering}
\includegraphics[height=5cm]{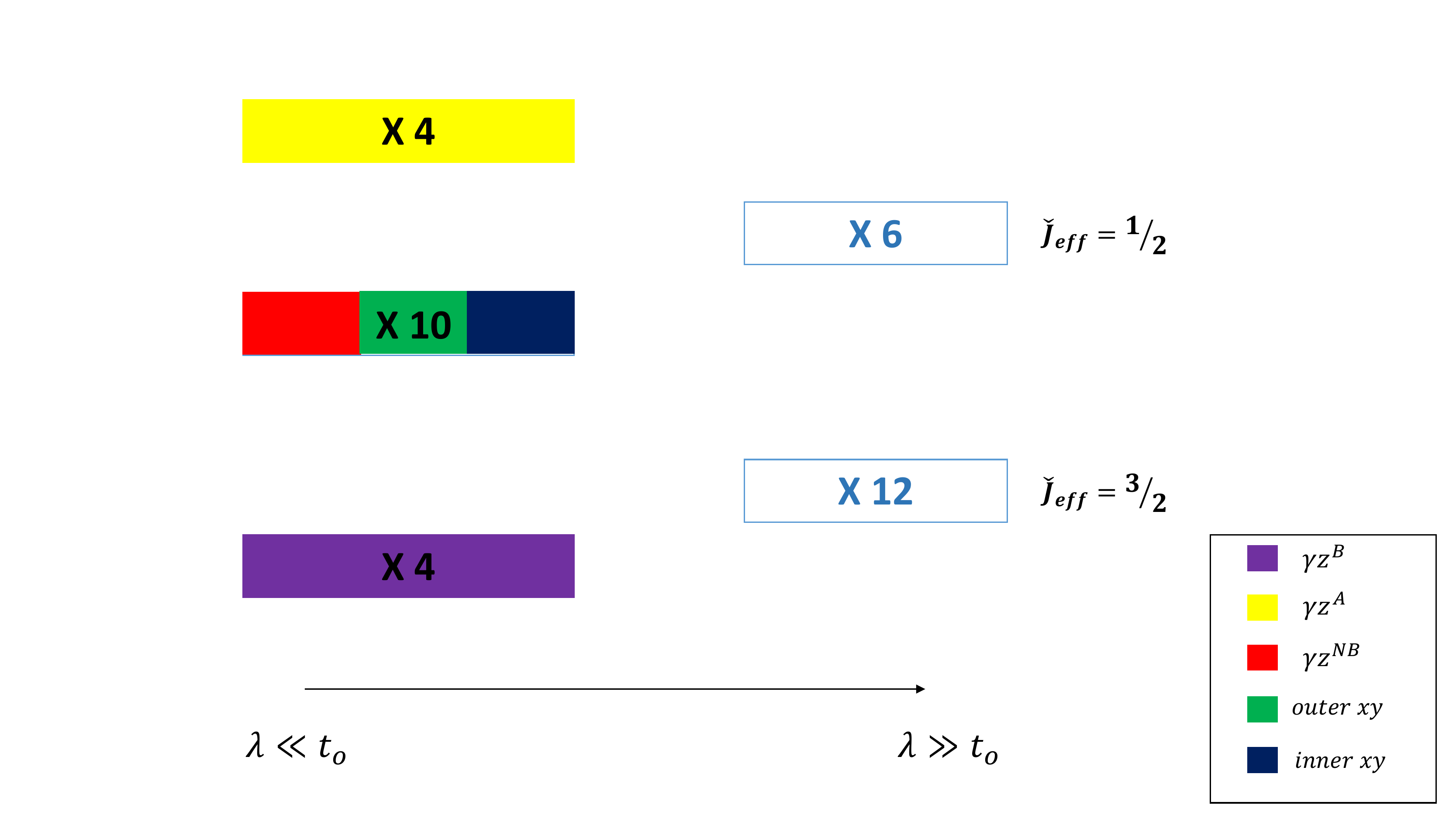}
\par\end{centering}
\caption{Schematic evolution of the energy levels of the unit cell Hamiltonian as functions of orthogonal hopping t$_o$ and SOC $\lambda$ without CF distortions. When $\lambda<<t_o$, three degenerate energy levels emerge, while in the limit of $\lambda>>t_o$, the system evolves into two blocks, constituted by two multiplets consisting of six and twelve levels respectively. The values of the effective total angular momentum $\tilde{J}_{eff}$, which are referred to each layer, are also indicated.}
\label{tolam}
\end{figure}

Let us start from the case of full local orbital degeneracy. The structure of the energy spectrum is made of three distinct eigenvalues as shown schematically in the Fig. \ref{tolam}.

\begin{figure}[h]
\noindent \begin{centering}
\includegraphics[height=5cm]{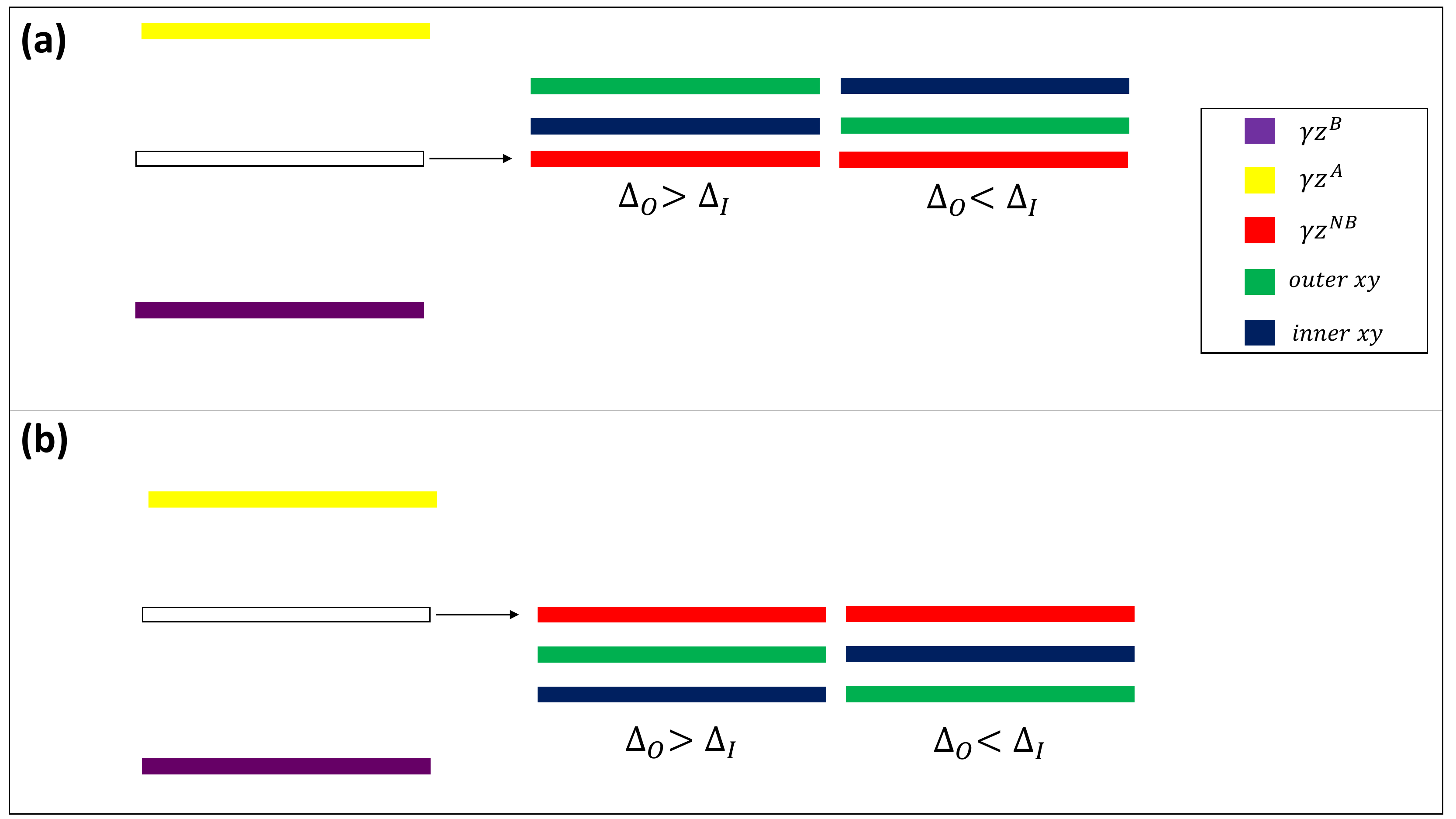}
\par\end{centering}
\caption{Schematic representation of the orbital configurations obtained by diagonalizing the Hamiltonian \eqref{HamiltTr} at $t_p$=0 with ${\Delta_O}$ and ${\Delta_I}$ concordant and positive (a) or negative (b).}
\label{conposneg}
\end{figure}

\begin{figure}[h]
\noindent \begin{centering}
\includegraphics[height=5cm]{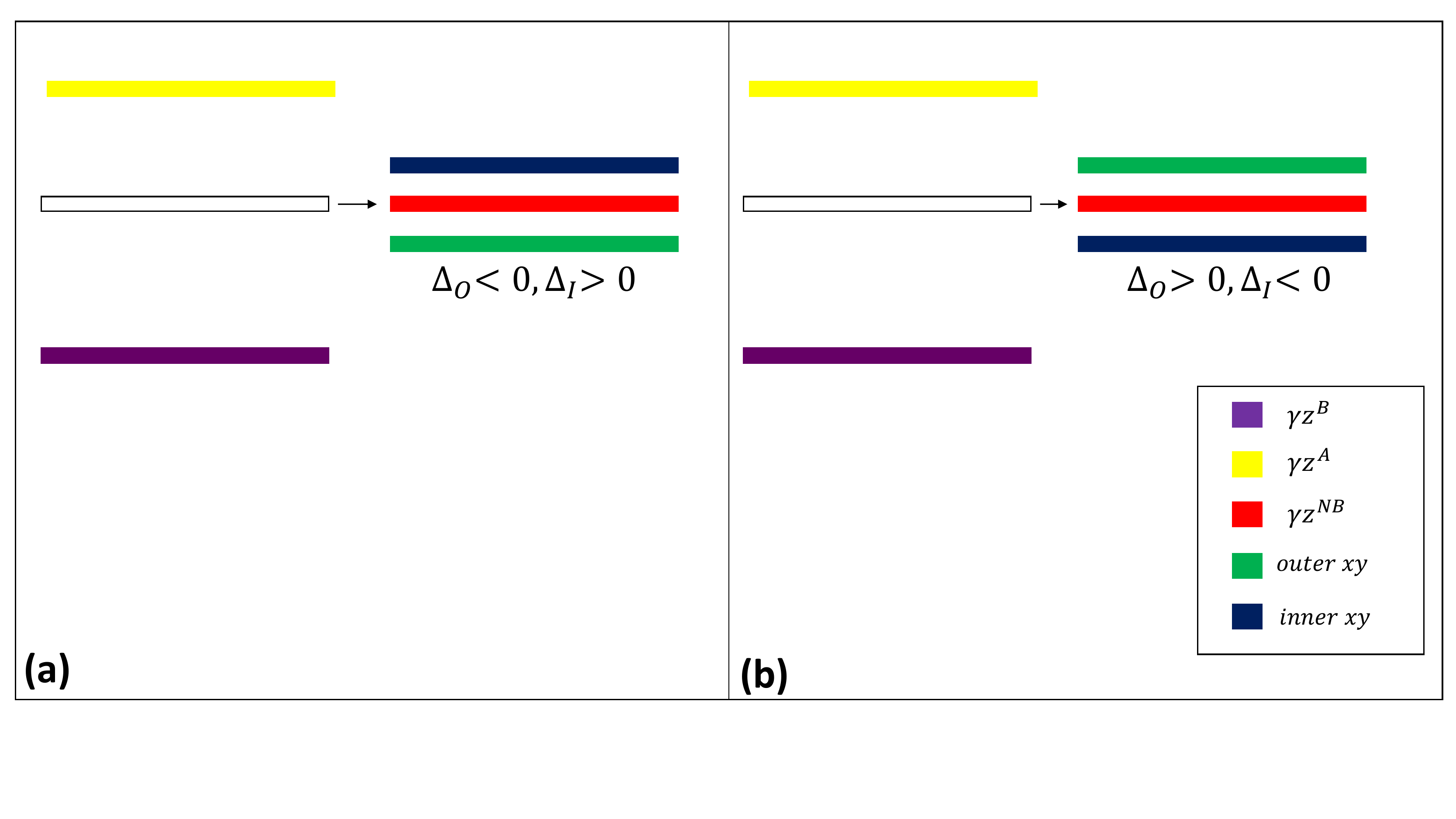}
\par\end{centering}
\caption{Schematic representation of the orbital configurations obtained by diagonalizing the Hamiltonian \eqref{HamiltTr} at $t_p$=0 with ${\Delta_O}$<0 and  ${\Delta_I}$>0 (a) or ${\Delta_O}$>0 and  ${\Delta_I}$<0 (b).   }
\label{disc}
\end{figure}

The lower block is constituted by the bonding $\gamma$z states $\gamma$z$^B$, while the upper block is composed by the antibonding $\gamma$z states $\gamma$z$^A$ (each of them with spin up and down); in the central block, there are non-bonding $\gamma$z states $\gamma$z$^{NB}$  and the six $xy$ local configurations.\\
By switching on the crystal field potential $\Delta_O$ and $\Delta_I$, one may simulate different local electronic environments and partially lift the orbital degeneracy. In particular, depending on the character of the structural distortions, apart from a drive in the amplitude, the sign attributed to $\Delta_O$ and $\Delta_I$ can be positive or negative. For instance, in a perovskite environment, if the t$_{2g}$ orbitals are in an octahedral cage, then a compression or elongation of the octahedra can lead to different types of level splitting between the $\gamma$z and $xy$ orbitals. Since we choose to measure the one-site orbital energy with respect to the $\gamma$z orbitals, positive (negative) values of ${\Delta_O}$ and ${\Delta_I}$ will raise (lower) the energy of the outer and inner $xy$ orbitals with respect to the $\gamma$z$^{NB}$. In Figs. \ref{conposneg} and \ref{disc}, we report the schematics of all the possible configurations that can be obtained by considering various choices of the crystal field potential terms ${\Delta_O}$ and ${\Delta_I}$.

\begin{figure}[h]
\noindent \begin{centering}
\includegraphics[height=5cm]{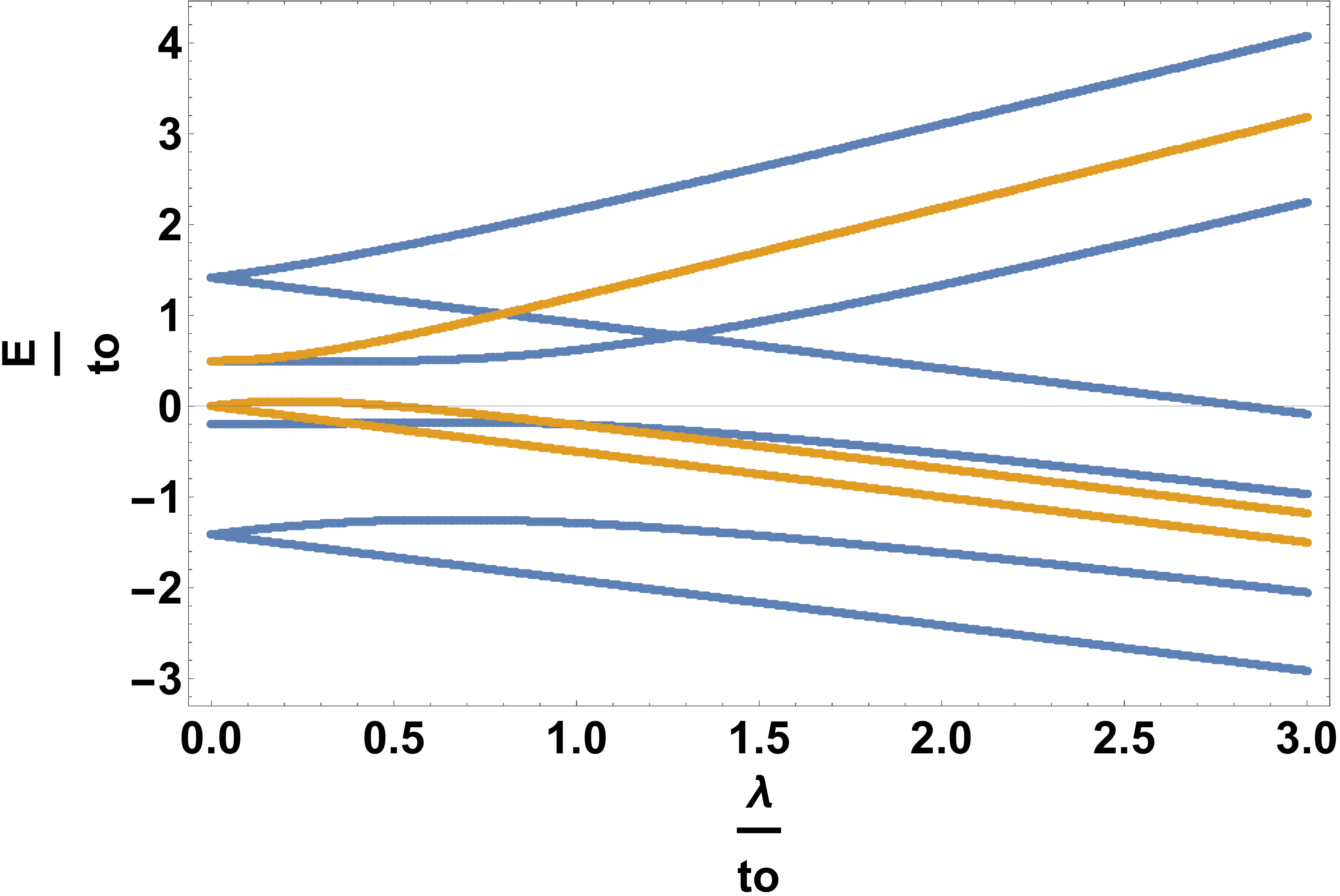}
\par\end{centering}
\caption{Evolution of the energy levels as function of $\lambda$. Blue energy levels are characterized by $\textit{f}$=1, while orange ones have $\textit{f}$=-1. All the parameters are expressed in unit of t$_o$. Here, $\frac{\Delta_I}{t_o}$=-0.2 and $\frac{\Delta_O}{t_o}$=0.5}
\label{socimopbis}
\end{figure}

\begin{figure}[h]
\noindent \begin{centering}
\includegraphics[height=5cm]{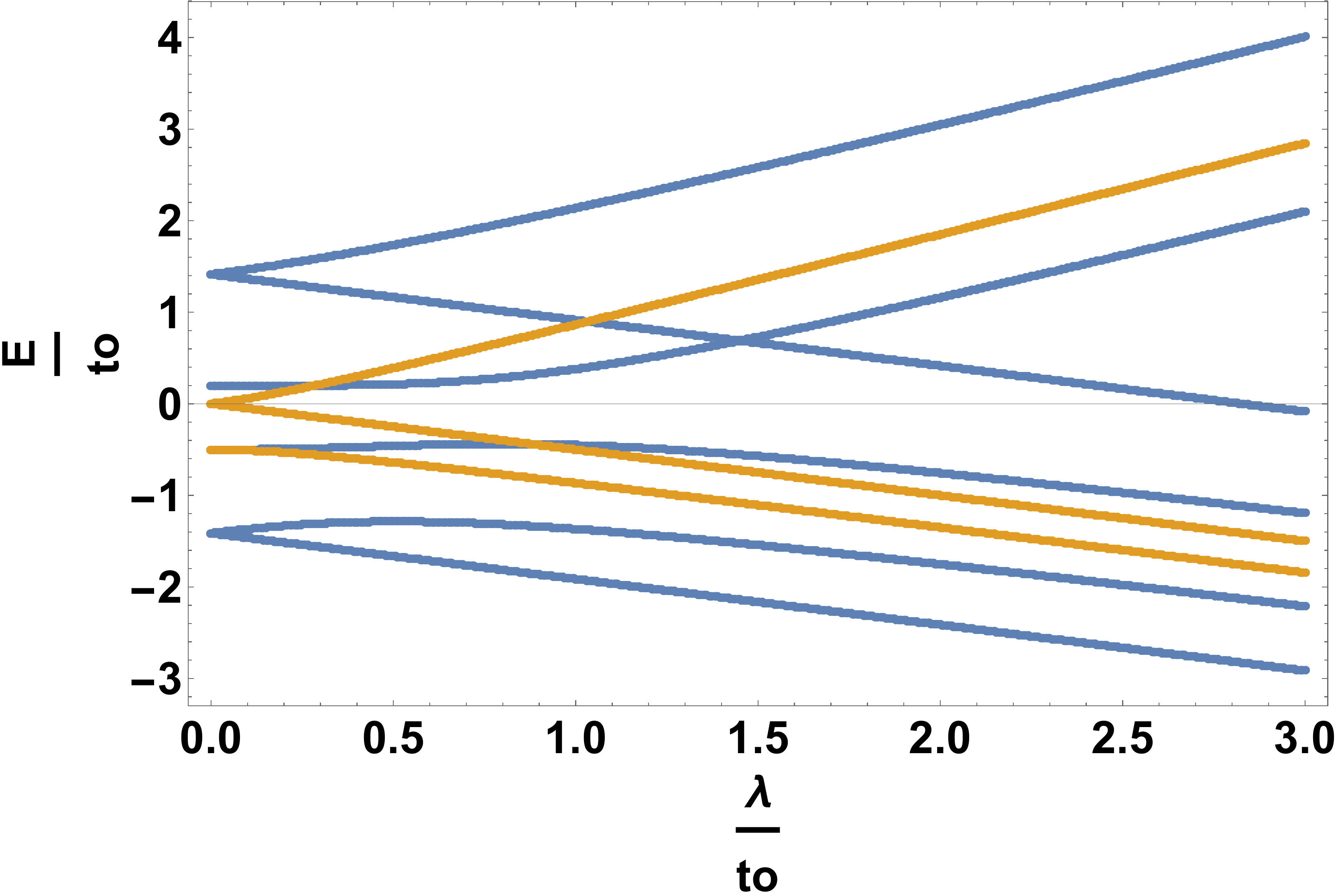}
\par\end{centering}
\caption{Evolution of the energy levels as function of $\lambda$.  Blue energy levels are characterized by $\textit{f}$=1, while orange ones have $\textit{f}$=-1. All the parameters are expressed in unit of t$_o$. Here, $\frac{\Delta_I}{t_o}$=0.2 and $\frac{\Delta_o}{t_o}$=-0.5.}
\label{socipombis}
\end{figure}

Then, we move further to discuss the consequences of a non-vanishing atomic spin-orbit in setting the structure of the local energy configurations. In particular, we consider two representative cases with $\Delta_I$ and $\Delta_O$ being unequal in amplitude and opposite in sign as depicted in Fig. \ref{disc} (a) and (b). For convenience all the parameters are expressed in units of t$_o$. Fig. \ref{socimopbis} and \ref{socipombis} show the evolution of the energy levels as function of $\lambda$ for the specific cases of $\frac{\Delta_I}{t_o}$=-0.2, $\frac{\Delta_O}{t_o}$=0.5 and $\frac{\Delta_I}{t_o}$=0.2, $\frac{\Delta_O}{t_o}$=-0.5. Each state is marked with a color which is representative of its layer parity value $\textit{f}$. As one can see, by adding the SOC, the intra-unit cell degeneracy which characterizes the local energy states at $\lambda$=0 is fully removed. However, the splitting and the hierarchy of the energy levels manifest distinct features for a given choice of the inner and outer CF, meaning that the interplay between these electronic parameters is nontrivial.

We also observe that for such choice of the layer dependent crystal field potentials, levels from the 3$^{rd}$ to the 8$^{th}$ in ascending order cross at degeneracy points when $\lambda$ is increased.
Levels with opposite parity generally cross at a critical $\lambda$, thus determining the subsequent exchange of $\textit{f}$ in the sequence of the parity of levels. We distinguish two regions: for small values of $\lambda$, Fig. \ref{socimopbis} shows a crossing between levels 3 and 4, while in Fig. \ref{socipombis} crossing between levels 6 and 7 arises; for moderate/large value of SOC, crossing is obtained for 4,5 and 7,8 levels in both cases. 

\section{Inversion layer symmetry protection of nodal loops and topological transitions from single to multiple nodal loops}

In this Section we discuss the electronic character of the nodal lines that arise from the crossings of bands having opposite layer parity. Due to the layer inversion symmetry bands with parity +1 and -1 cannot hybridize, therefore can cross each other at $k$-points satisfying the relation
\begin{equation}
E_{+}(k)=E_{-}(k) \,.
\end{equation} 

Since $k$ has two free components (we deal with a 2D system), we have two variables that have to satisfy one equation, meaning that the solution space is generically one-dimensional, i.e. a nodal line. Since the nodal line is confined in a plane, one can use zero dimensional manifolds to characterize its topological structure. In particular, we employ a criterion which allows to directly identify the presence of bands crossing and the possibility of a transition in the structure of the nodal line. We fix the chemical potential in a way that only a pair of bands is involved and we determine the parity of the band below the Fermi level in each high symmetry position of the Brillouin zone. If the parity is not the same between the two symmetry positions, we associate a value of the parity number $\mathscr{I}=1$, then one can state that there will be at least one crossing along the $k$-direction connecting the two high symmetry points. Otherwise, there can be zero or an even number of band crossings if no change of parity is observed, $\mathscr{I}=0$.

Starting from the local energy analysis, we consider a representative electron filling with 4th occupied bands (i.e. eight electrons in the unit cell due to the time-inversion symmetry) and we evaluate $\mathscr{I}$ by comparing the band parity at the high symmetry points in the Brillouin zone, i.e. $\Gamma=[0,0]$, $X=[\pi,0]$, and $M=[\pi,\pi]$. Hence, $\mathscr{I}{\Gamma X}$ will provide the parity difference at the $\Gamma$ and $X$ points in the Briloouin zone and, thus, indicate the presence of even (including zero) or odd number of crossings along the $\Gamma$-$X$ direction.

\begin{figure}[t]
\noindent \begin{centering}
\includegraphics[height=5cm]{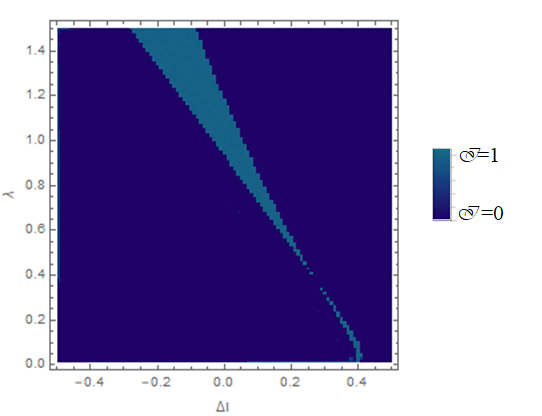}\hspace*{0.5cm}
\includegraphics[height=2cm]{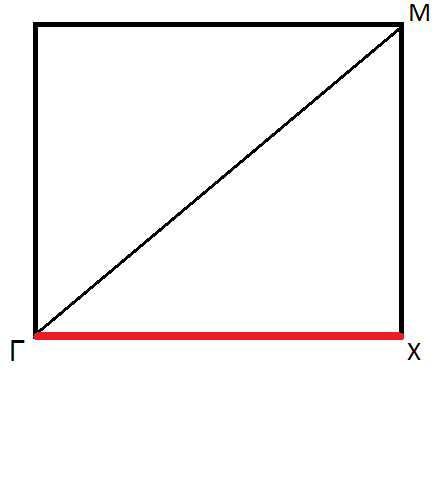}
\includegraphics[height=5cm]{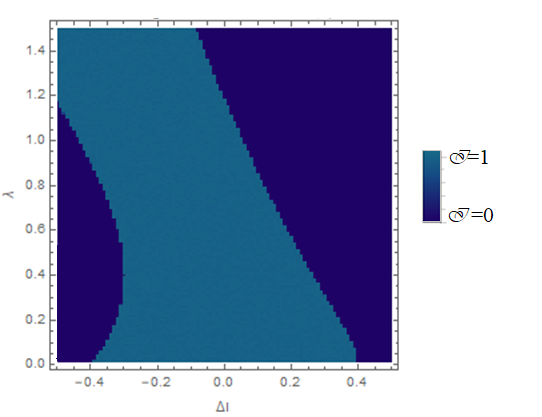}\hspace*{0.5cm}
\includegraphics[height=2cm]{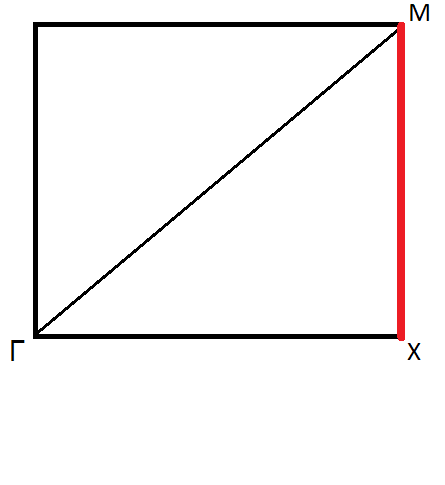}
\includegraphics[height=5cm]{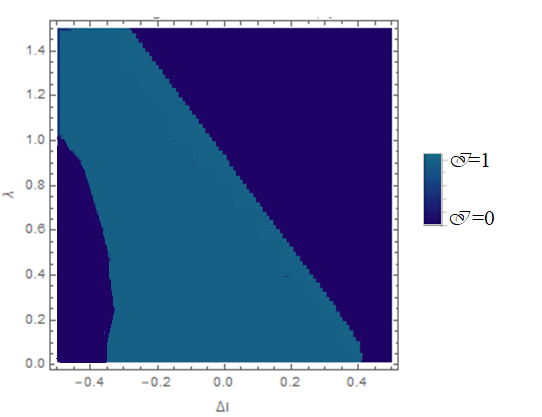}\hspace*{0.5cm}
\includegraphics[height=2cm]{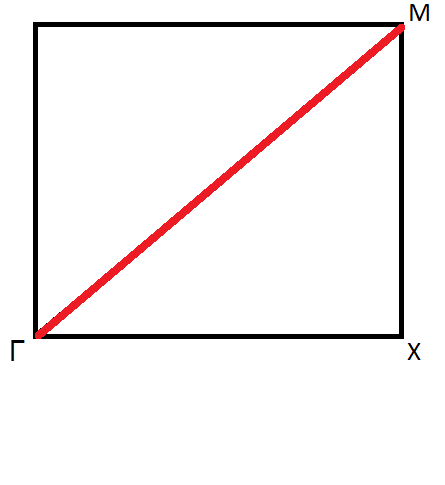}
\par\end{centering}
\caption{Phase diagram of the parity number $\mathscr{I}$ associated to the various directions in the Brillouin zone, as indicated by the red line in the figure on the right, for $\frac{\Delta_O}{t_o}$=0.5 and $\frac{t_p}{t_o}$=0.1. The dark blue zone corresponds to the value 0 of $\mathscr{I}$, while the light blue zone is relative to $\mathscr{I}$=1.}
\label{pdb4b505gx}
\end{figure}

\begin{figure}[t]
\noindent \begin{centering}
\includegraphics[height=5cm]{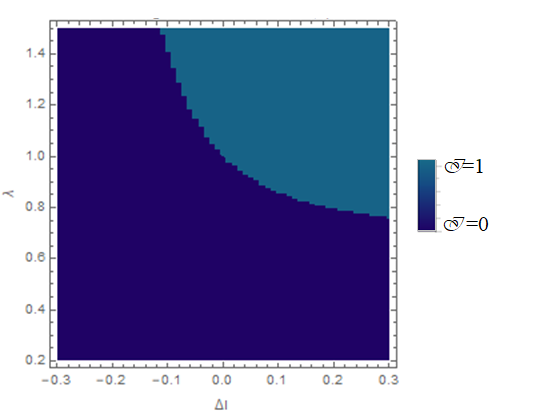}\hspace*{0.5cm}
\includegraphics[height=2cm]{gx}
\includegraphics[height=5cm]{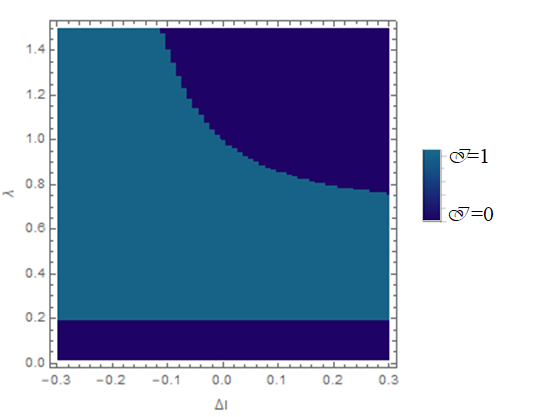}\hspace*{0.5cm}
\includegraphics[height=2cm]{xm}
\includegraphics[height=5cm]{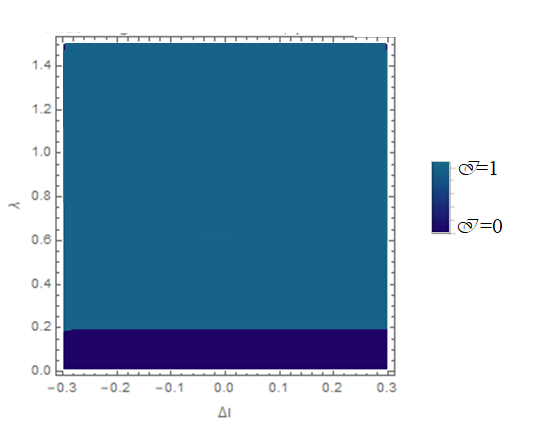}\hspace*{0.5cm}
\includegraphics[height=2cm]{gm}
\par\end{centering}
\caption{Phase diagram of the parity number $\mathscr{I}$ associated to the various directions in the Brillouin zone, as indicated by the red line in the figure on the right, for $\frac{\Delta_O}{t_o}$=-0.5 and $\frac{t_p}{t_o}$=0.1. The dark blue zone corresponds to the value 0 of $\mathscr{I}$, while the light blue zone refers to $\mathscr{I}$=1.}
\label{pdb4b5m05gxxm}
\end{figure}

Figs. \ref{pdb4b505gx} report the phase diagrams relative to a representative filling of 4 bands along $\Gamma$X, XM and $\Gamma$M directions, respectively, for the specific case of $\frac{\Delta_O}{t_o}$=0.5 and $\frac{t_p}{t_o}$=0.1. Looking at the phase diagrams, we can recognize two different parity regions: in one of them, the parity number is equal to zero, in the other it is equal to 1.\\
The value of the parameters is relevant in determining the structure of the phase diagram. Indeed, if we change the value of $\frac{\Delta_O}{t_o}$=-0.5, the phase diagrams which we obtain for the same filling are significantly modified (see Figs. \ref{pdb4b5m05gxxm}).

\begin{figure}[t]
\noindent \begin{centering}
\includegraphics[height=2cm]{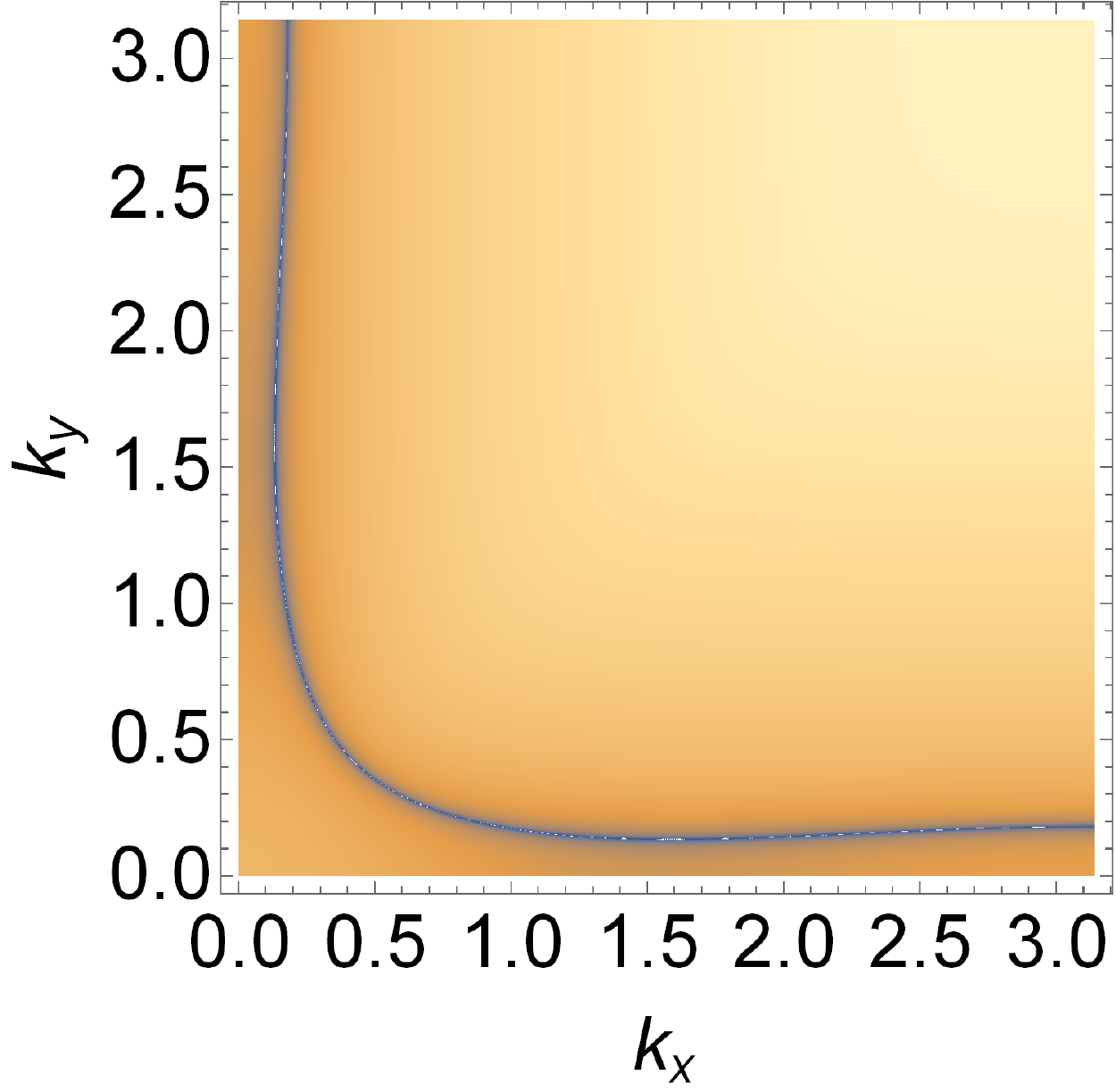}\hspace*{0.1cm}
\includegraphics[height=2cm]{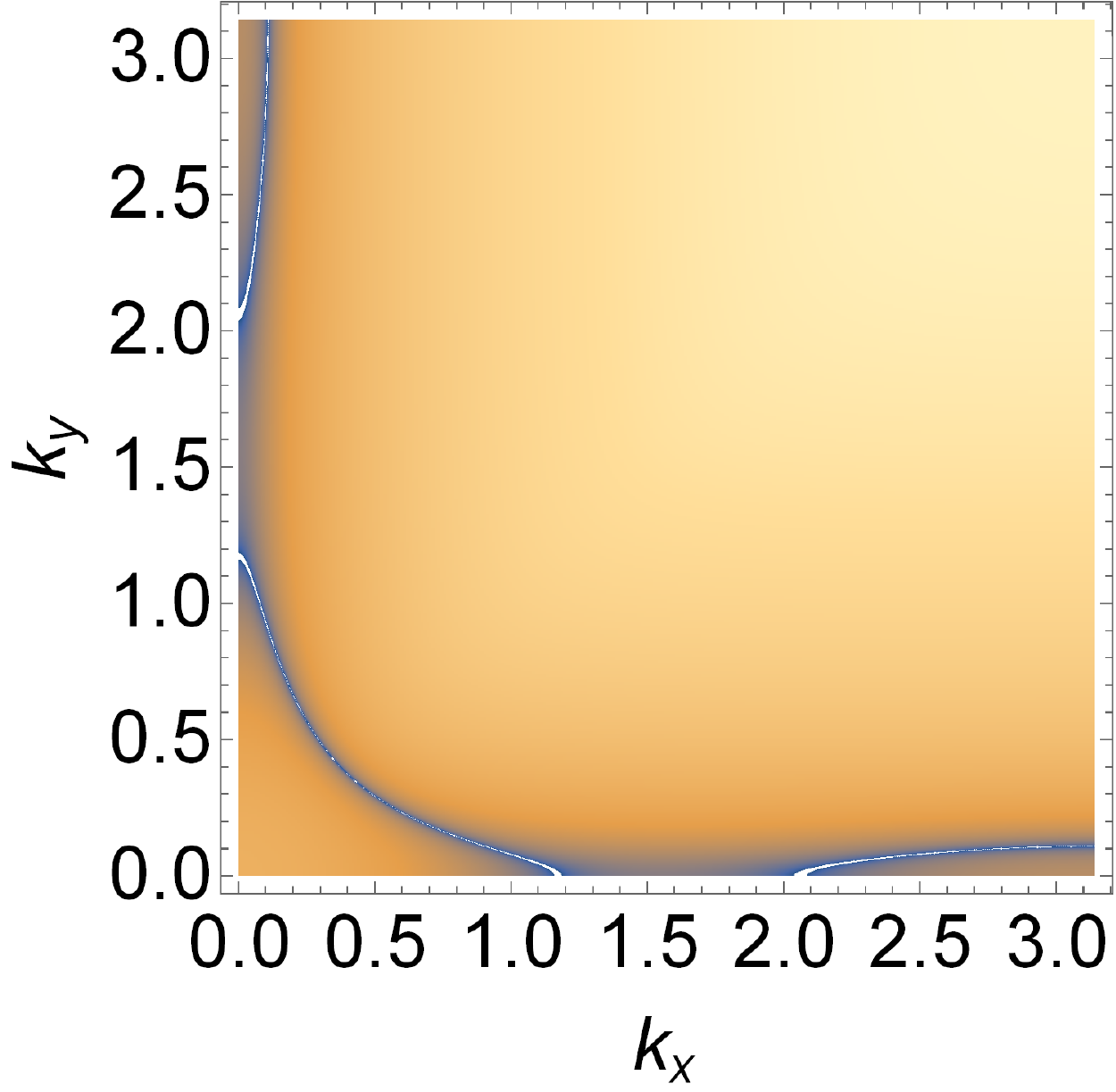}\hspace*{0.1cm}
\includegraphics[height=2cm]{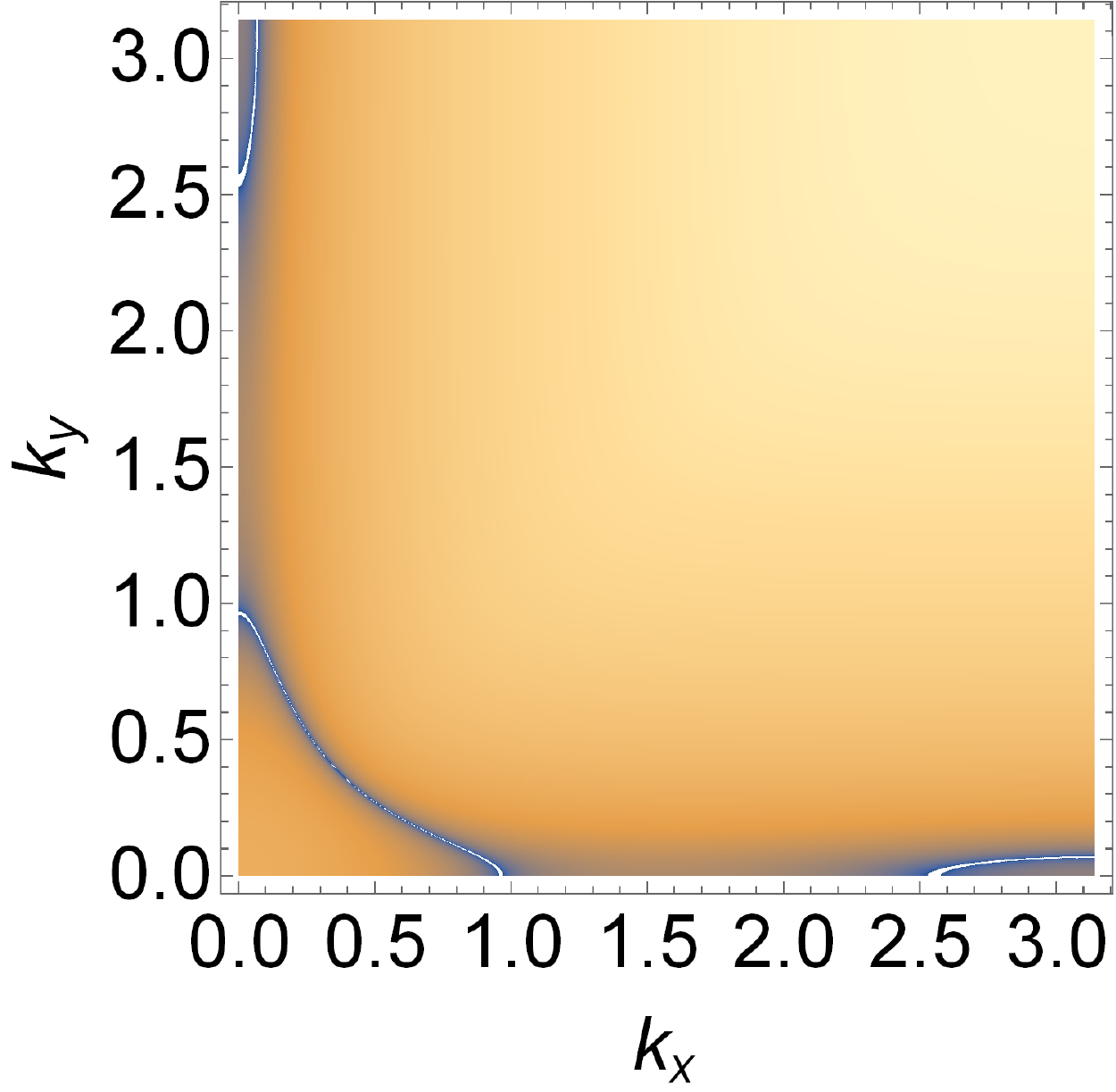}\hspace*{0.1cm}
\includegraphics[height=2cm]{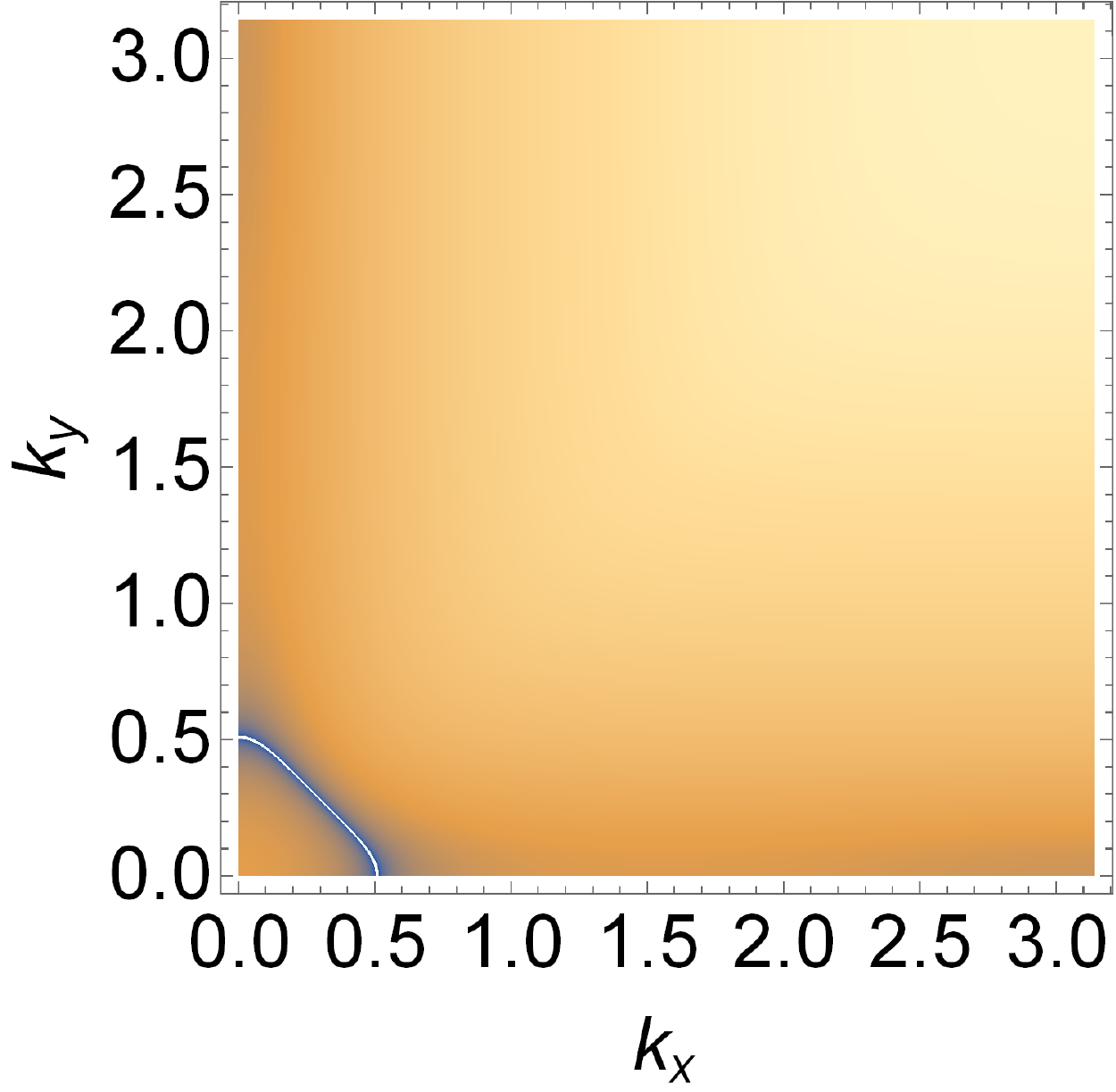}\hspace*{0.1cm}
\par\end{centering}
\caption{Nodal line transitions in momentum space for $\frac{t_p}{t_o}$=0.1,$\frac{\Delta_O}{t_o}$=0.5 . The gray line indicates the position of the crossing between the band 4 and 5 increasing the value of $\Delta_I$ for $\frac{\lambda}{t_o}$=0.2. From the left to the right: $\frac{\Delta_I}{t_o}$=0.34, $\frac{\Delta_I}{t_o}$=0.343, $\frac{\Delta_I}{t_o}$=0.344, $\frac{\Delta_I}{t_o}$=0.35}
\label{tb4b505bis}
\end{figure}

Remarkably, by varying the amplitude of $\lambda$ and $\Delta_I$ in the phase diagrams, one can achieve distinct electronic transitions between phases characterized by different topology of the nodal line loops. The character of each nodal structure at a given representative filling, involving a pair of crossing bands, is uniquely associated with the value of $\mathscr{I}$ computed at the high-symmetry points  ($\mathscr{I}_{\Gamma X}$,$\mathscr{I}_{XM}$,$\mathscr{I}_{\Gamma M}$). Here, we investigate several kinds of symmetry preserving transitions through which the nodal lines may evolve, when moving across different regions of the phase diagrams. We will consider both weak and strong SOC regimes if compared to the intra-unit cell electronic parameters.\\
We start by considering the phase diagrams reported in Figs.\ref{pdb4b505gx}. For $\frac{\lambda}{t_o}$=0.2, we vary $\Delta_I$ in the interval 0.34$\le\frac{\Delta_I}{t_o}\le$0.35. For this parameters scan, we can distinguish three distinct kinds of transition: $\mathscr{I}_{\Gamma X}$ goes from 0 to 1, $\mathscr{I}_{XM}$ goes from 1 to 0 while $\mathscr{I}_{\Gamma M}$ remains unchanged and equal to 1. We indicate such transition as 011 $\rightarrow$ 101, where each of the three numbers represent the value of the parity number along the $\Gamma$X, XM and $\Gamma$M direction, respectively. One can then follow the evolution of the nodal line associated to each band crossing in the chosen region of parameters. The outcome of this analysis is summarized in Fig. \ref{tb4b505bis}.

\begin{figure}[t]
\noindent \begin{centering}
\includegraphics[height=2cm]{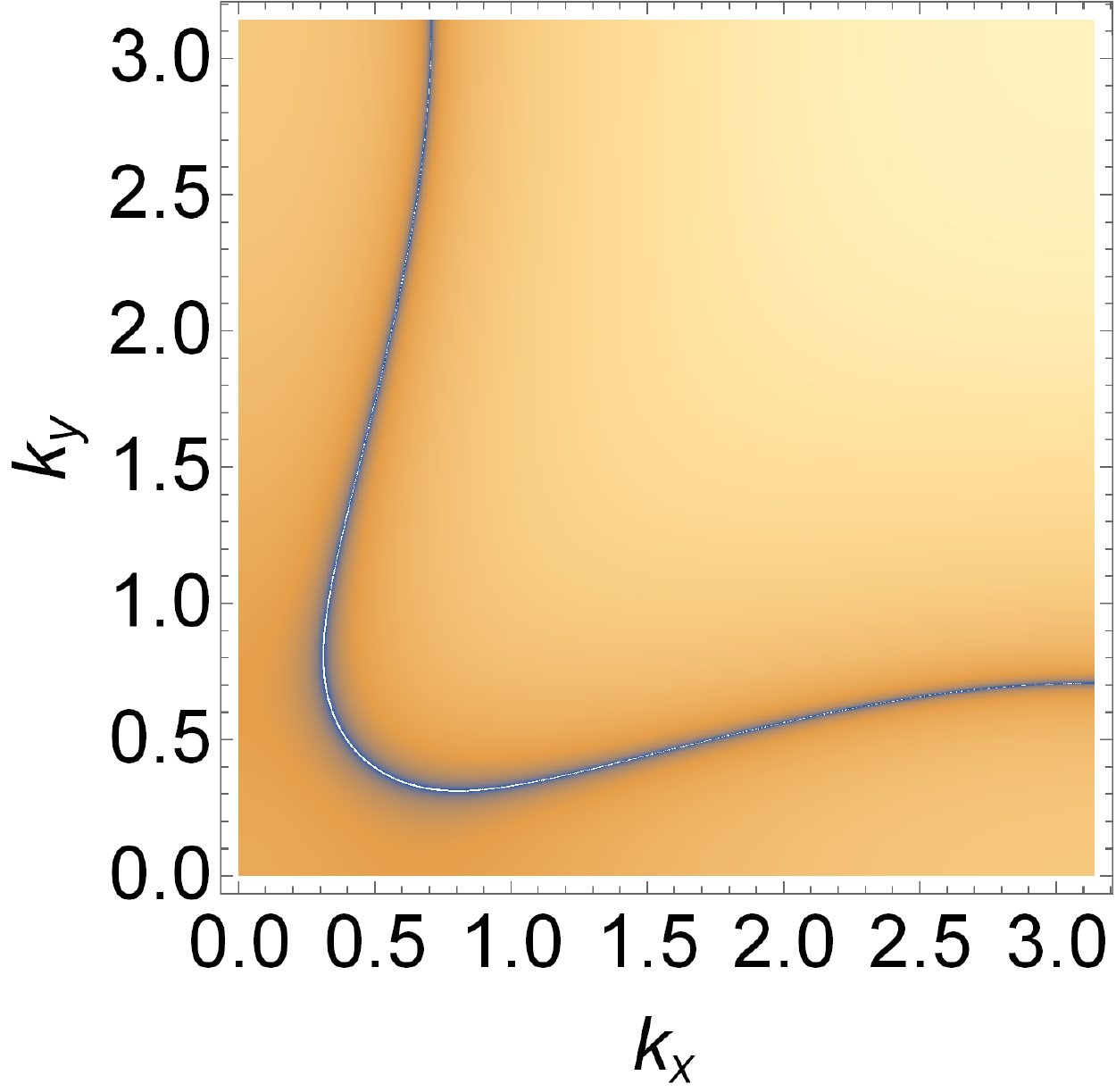}\hspace*{0.1cm}
\includegraphics[height=2cm]{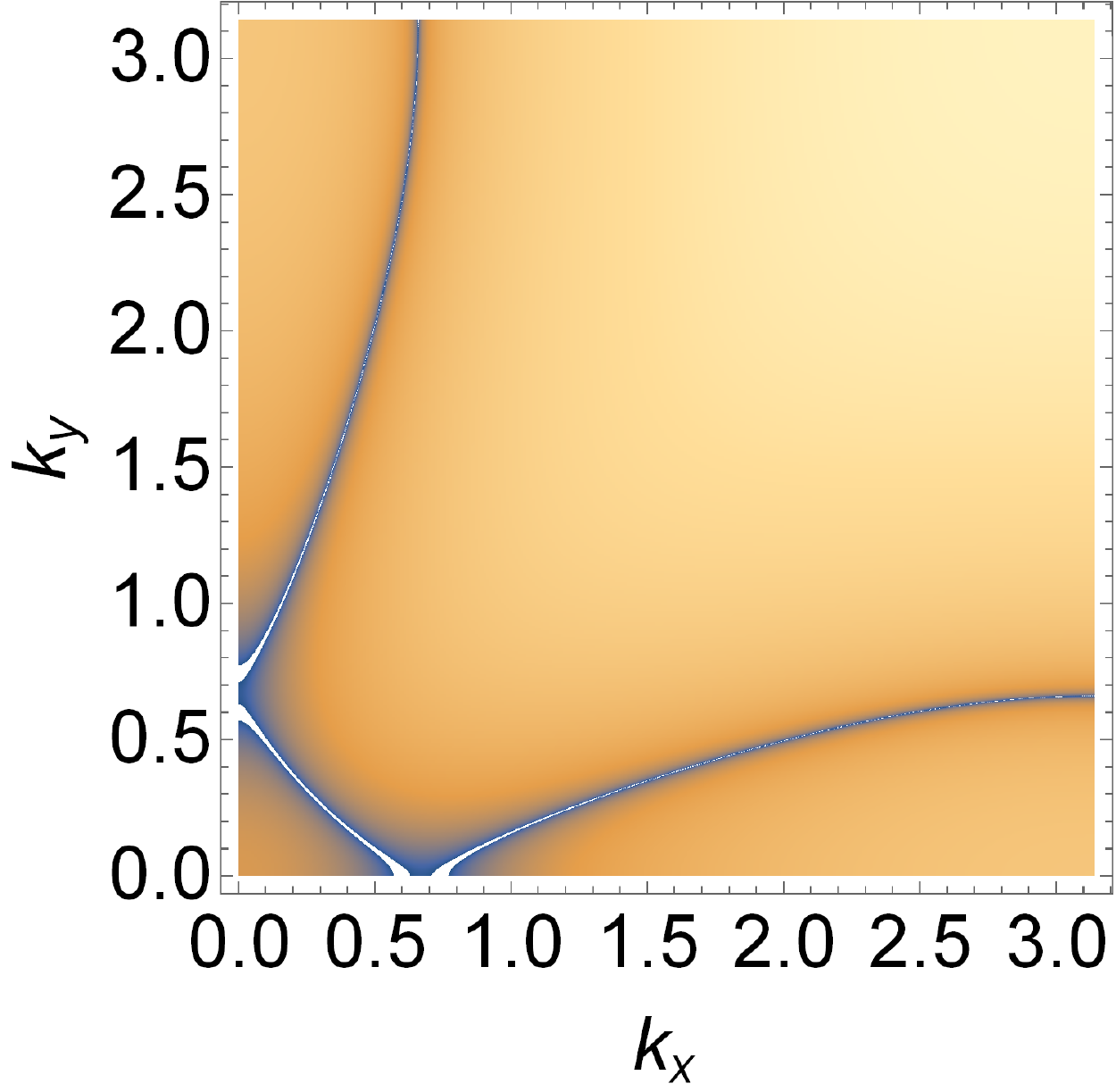}\hspace*{0.1cm}
\includegraphics[height=2cm]{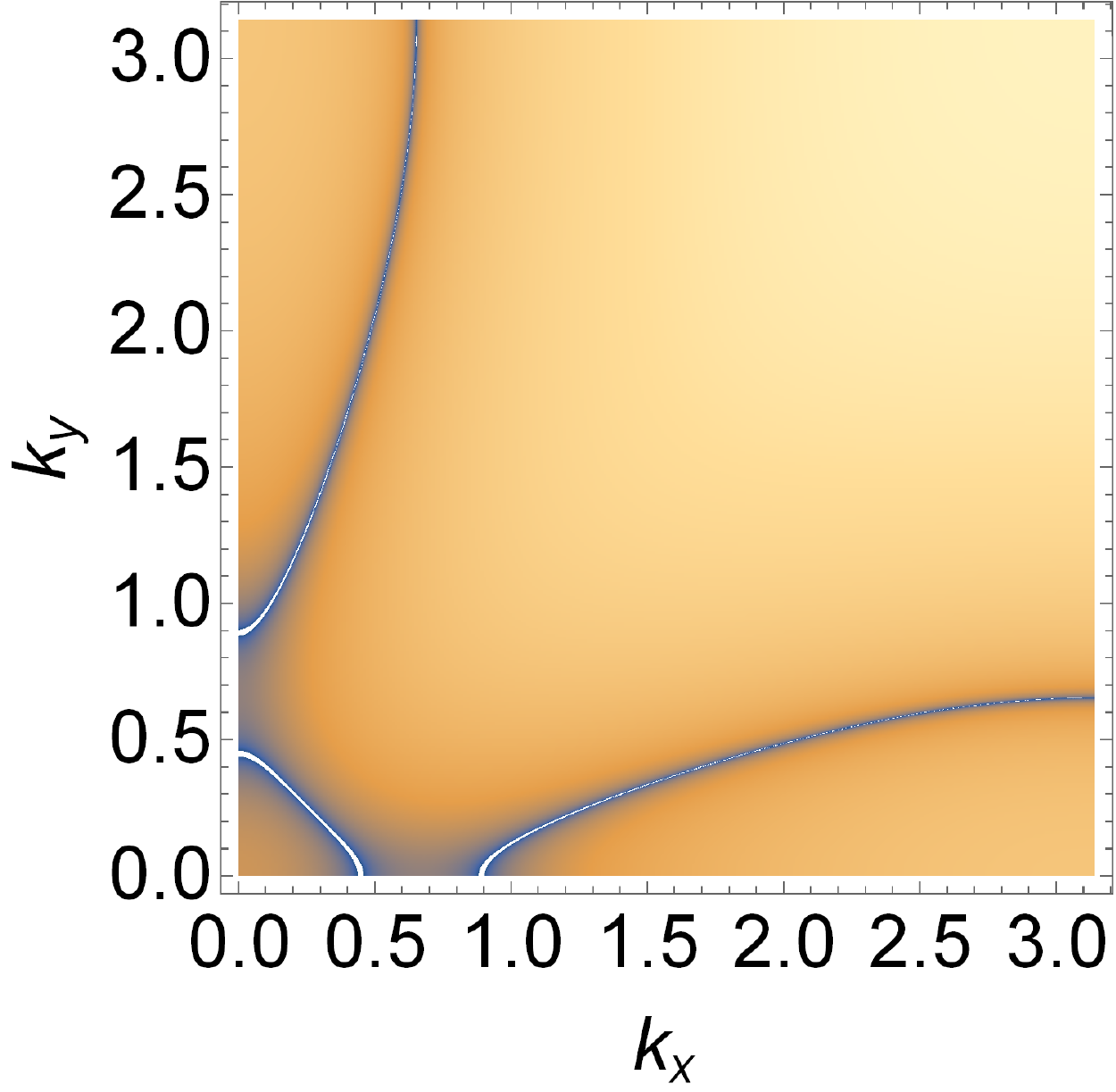}\hspace*{0.1cm}
\includegraphics[height=2cm]{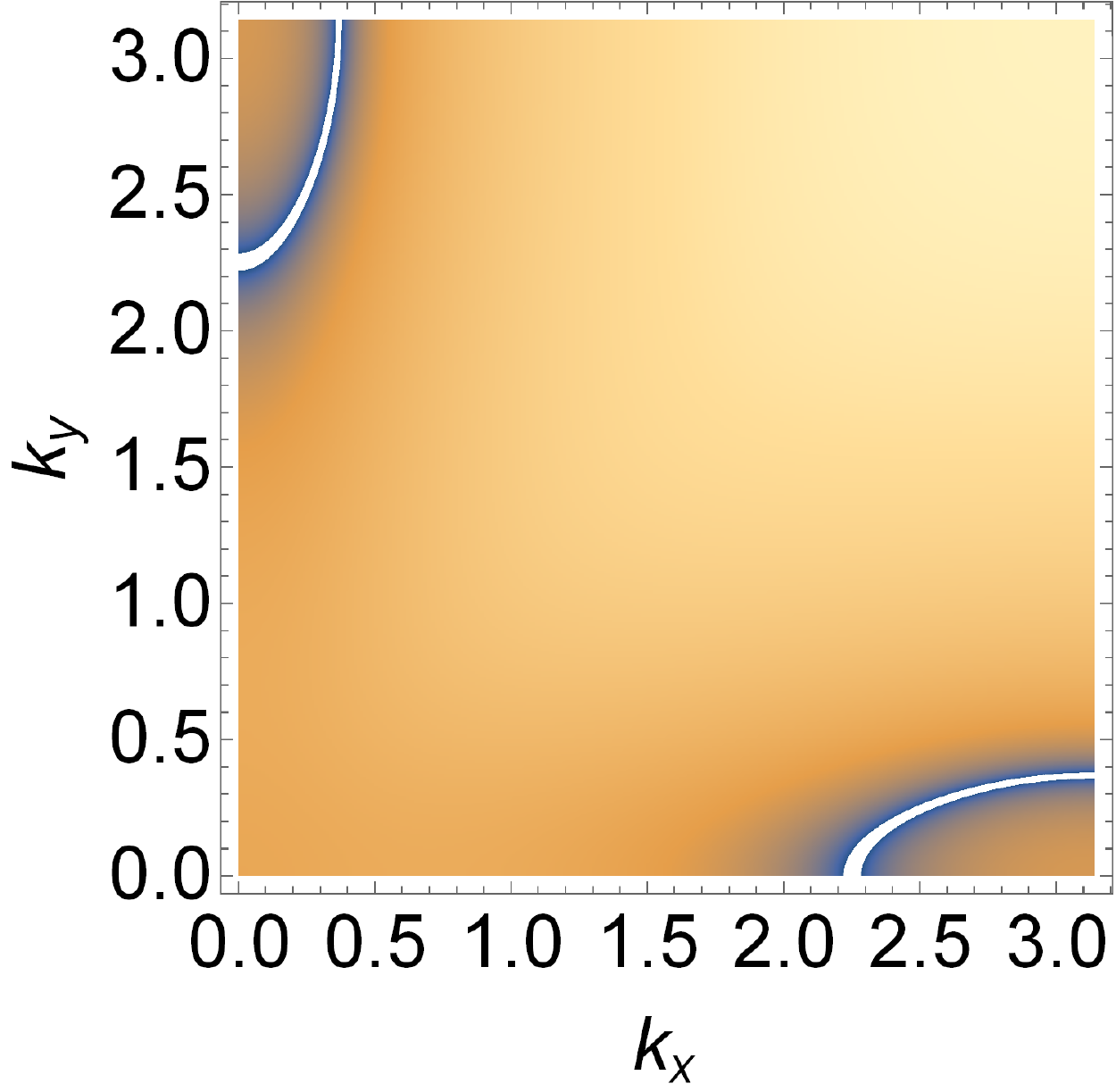}\hspace*{0.1cm}
\par\end{centering}
\caption{Nodal line transitions in the momentum space relative to the bands 4 and 5 with $\frac{t_p}{t_o}$=0.1,$\frac{\Delta_O}{t_o}$=0.5 . The gray line indicates the position of the crossing between the band 4 and 5 increasing the value of $\Delta_I$ for $\frac{\lambda}{t_o}$=0.8. From left to right: $\frac{\Delta_I}{t_o}$=0.05, $\frac{\Delta_I}{t_o}$=0.0588, $\frac{\Delta_I}{t_o}$=0.06, and $\frac{\Delta_I}{t_o}$=0.1.}
\label{tb4b505}
\end{figure}

In Fig. \ref{tb4b505bis}, the first panel on the left corresponds to $\frac{\Delta_I}{t_o}$=0.34 and the nodal line intercepts the XM and the $\Gamma$M direction, which means that the bands 4 and 5 exhibit a crossing point as due to a parity inversion along those directions. This is coherent with the mapping 011 which we extract from the phase diagrams of Fig. \ref{pdb4b505gx}. In the second panel, $\frac{\Delta_I}{t_o}$=0.343 and the topology of the line has changed: the nodal line collapses in the $\Gamma$X direction. This means that, along $\Gamma$ X, band 4 and 5 have two subsequent intersections, that is the reason why the total number of inversion, namely $\mathscr{I}_{\Gamma X}$, 
is equal to 0. The pocket around the X point, which is appearing in this range of parameters, disappears when the value of $\frac{\Delta_I}{t_o}$ is increased. In the last panel on the right, $\frac{\Delta_I}{t_o}$=0.35 and there is only one pocket around the $\Gamma$ point. This reproduces correctly the final values of the inversion parity numbers as 101, where the bands have no more intersections along XM direction.
In summary, when moving across such kind of transitions, a nodal line ( first panel on the left of Fig. \ref{tb4b505bis}) separates in two distinct (central panels of Fig. \ref{tb4b505bis}) nodal lines by collapsing on the k$_x$ axis. During the transition, the $\mathscr{I}_{\Gamma X}$ is initially conserved, until one nodal line disappears and $\mathscr{I}_{\Gamma X} \rightarrow$1. The transition has a topological nature because it is characterized by a modification of the topology of the nodal loops.\\
A similar transition can be obtained in the limit of strong SOC for the same electron filling, i.e. pair of bands. We consider the phase diagrams of Figs. \ref{pdb4b505gx} and, by fixing the value of $\lambda$ to $\frac{\lambda}{t_o}$=0.8, one can find a modification of the shape of the nodal loops for 0.05$\le\frac{\Delta_I}{t_0}\le$0.1 which exhibits a changeover of the type 011$\rightarrow$110. In that transition, shown in Fig. \ref{tb4b505}, a starting pocket around the M point collapses along the $\Gamma$X direction and subsequently disappears by closing itself around the $\Gamma$ point, while another loop around the X point gets formed.

\begin{figure}[t]
\noindent \begin{centering}
\hspace*{-1cm}
\includegraphics[height=2cm]{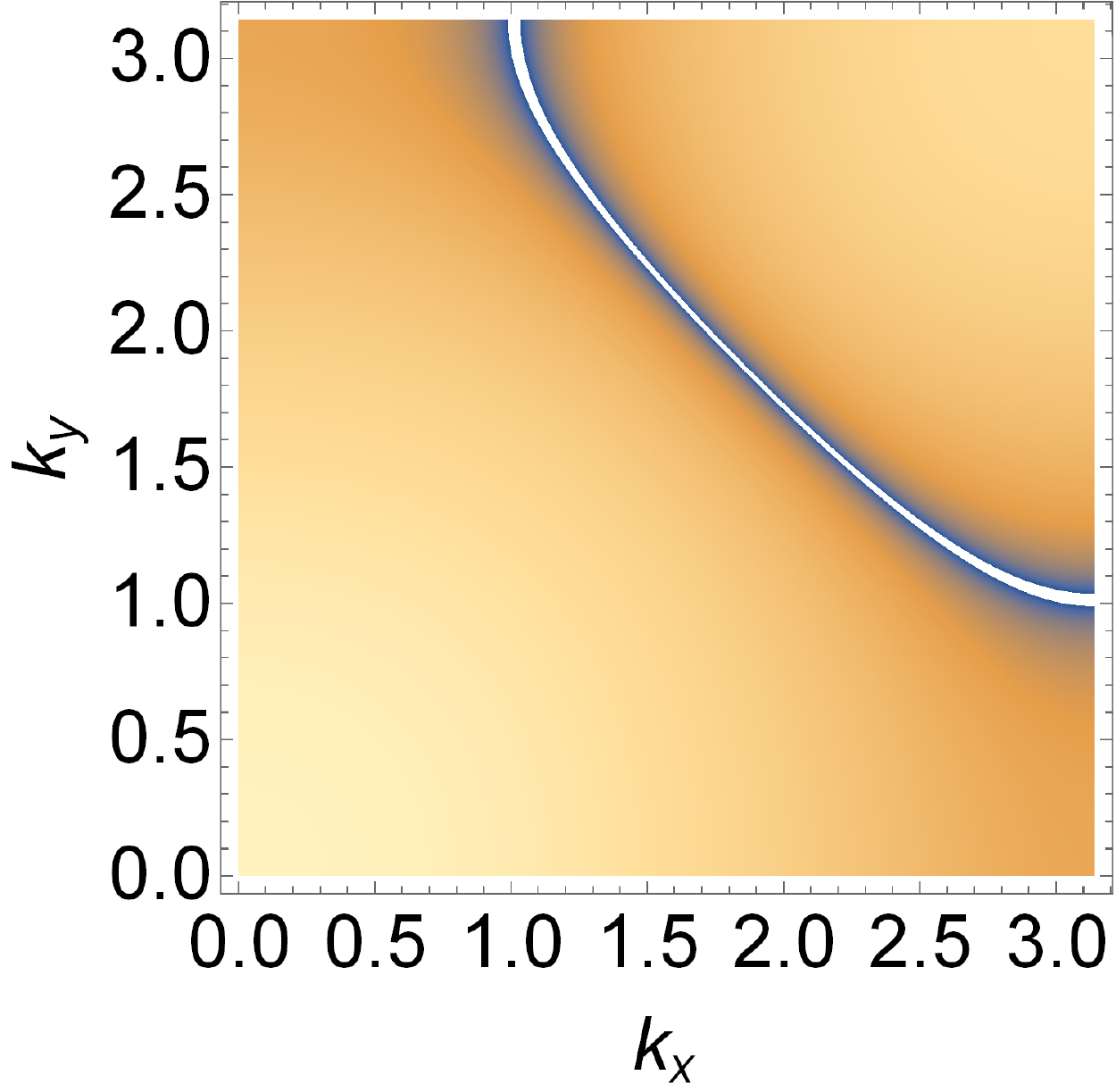}\hspace*{0.1cm}
\includegraphics[height=2cm]{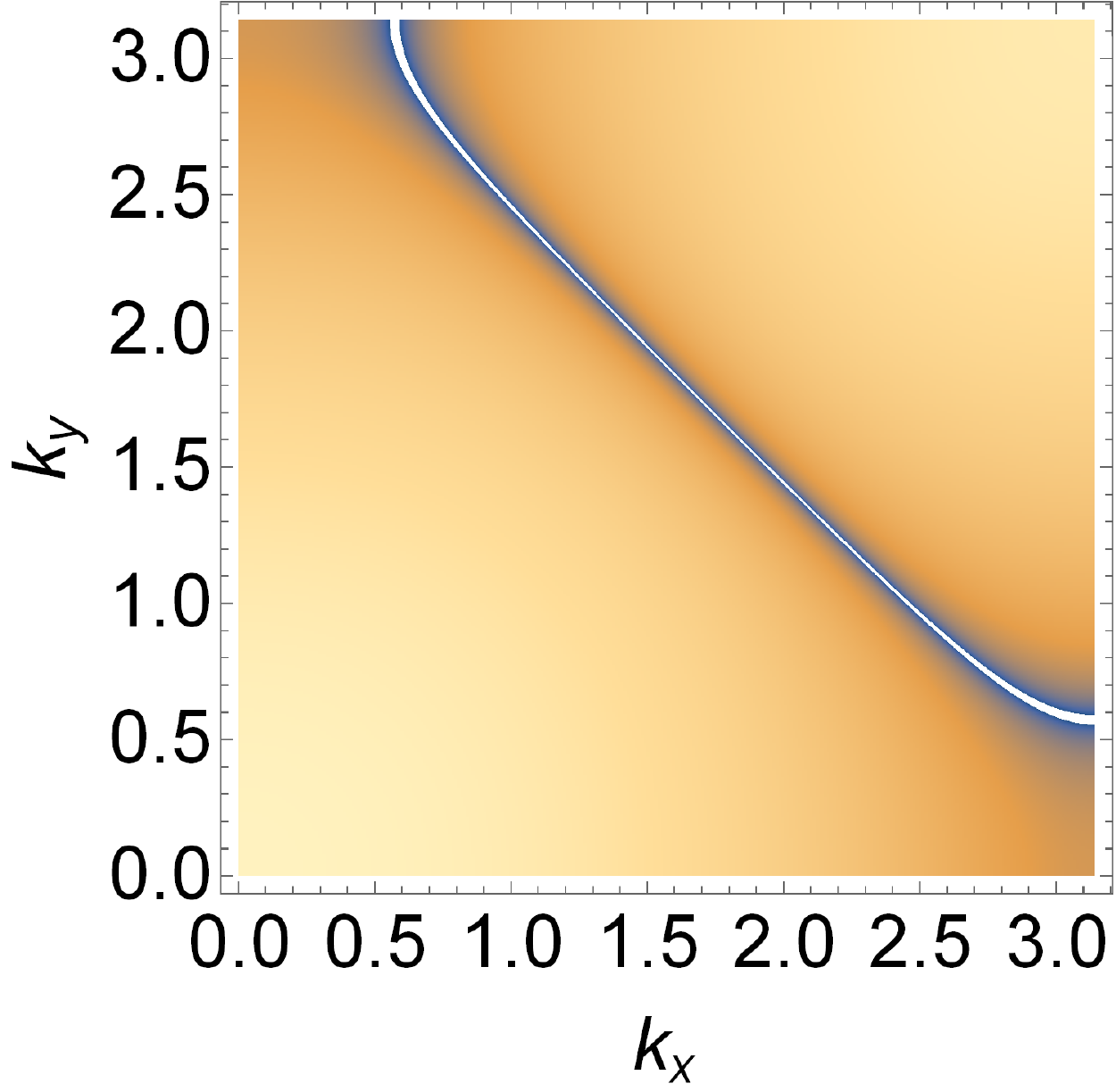}\hspace*{0.1cm}
\includegraphics[height=2cm]{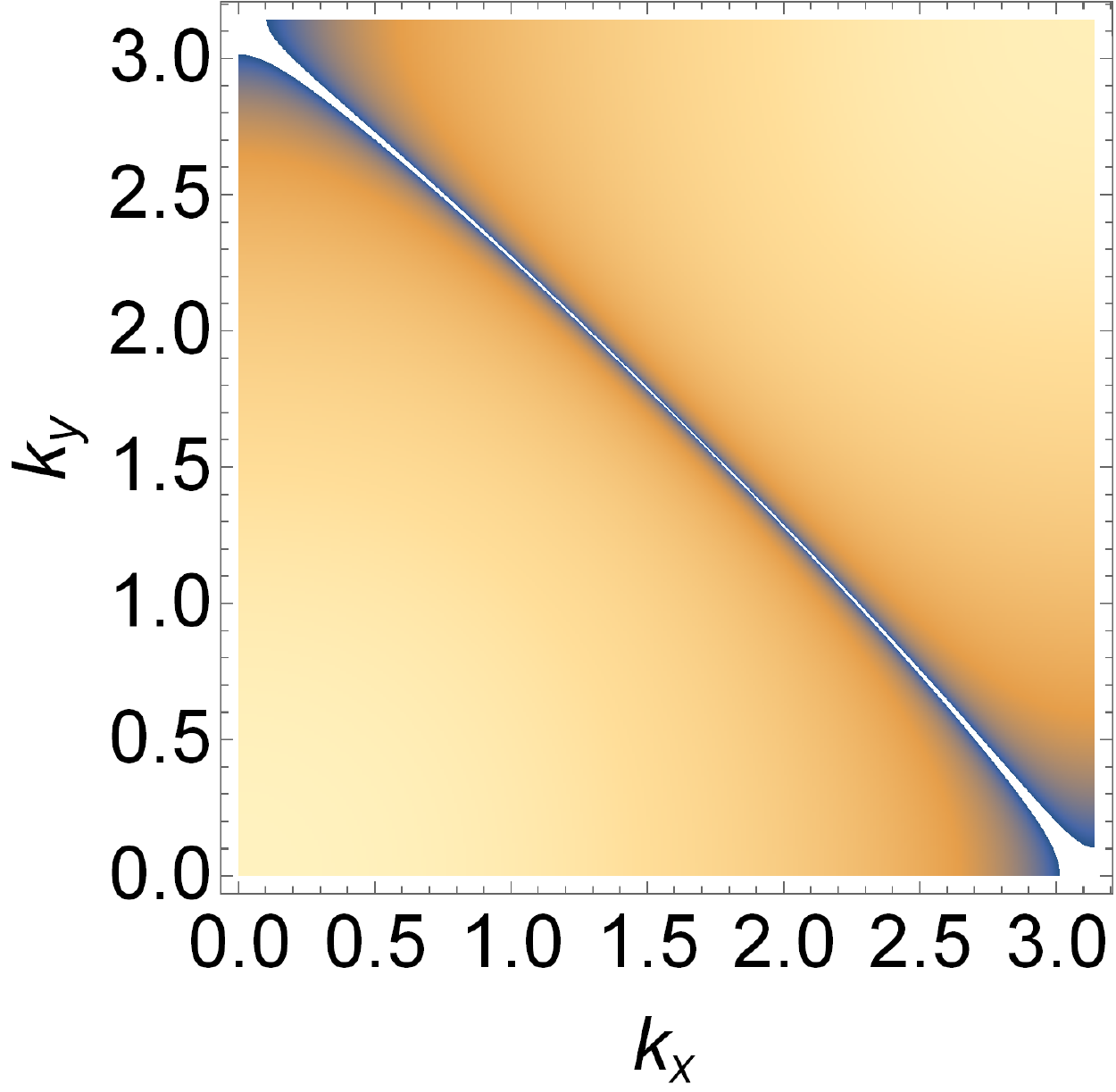}\hspace*{0.1cm}
\includegraphics[height=2cm]{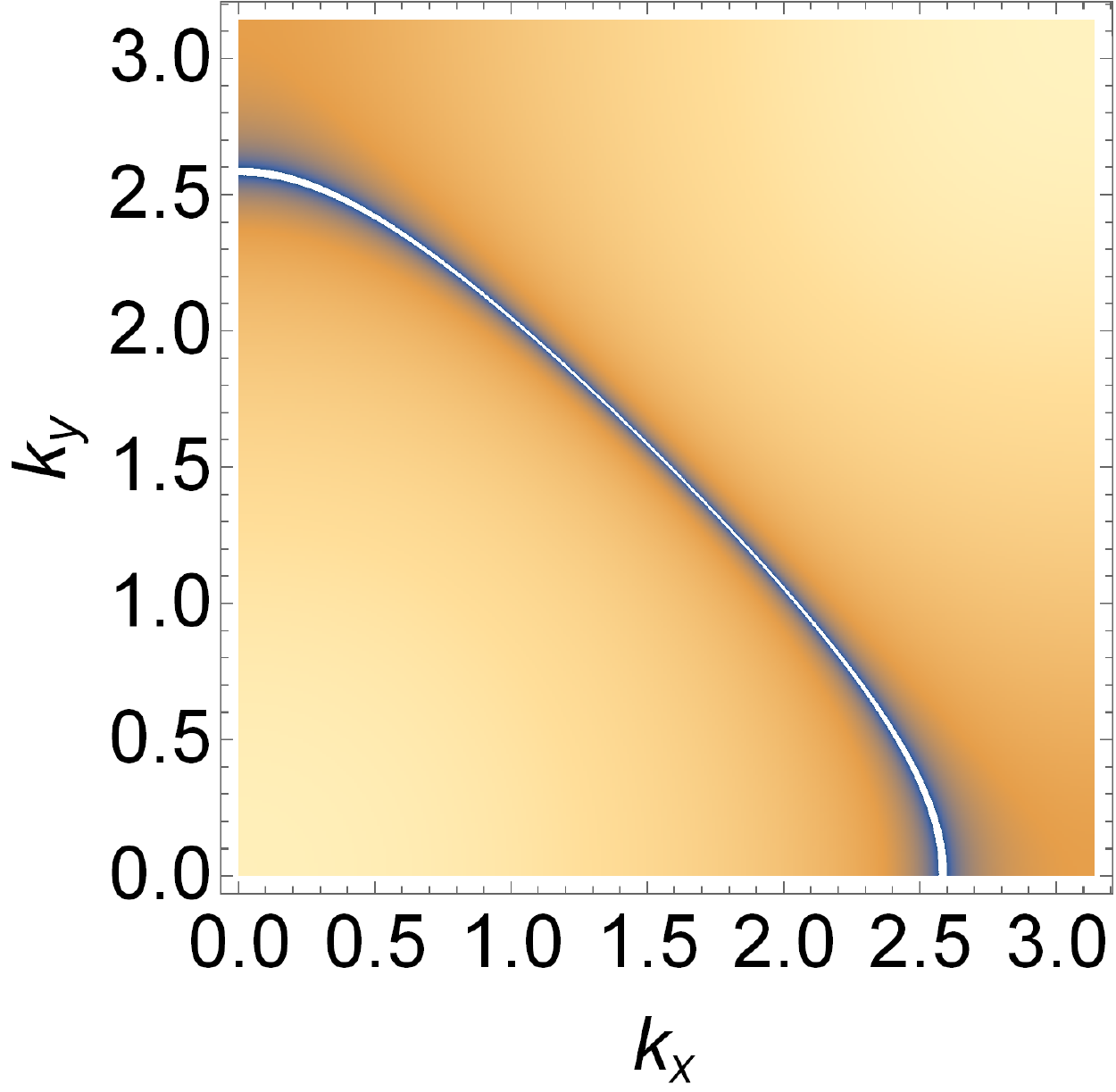}\hspace*{0.1cm}
\par\end{centering}
\caption{Nodal line transitions in the momentum space relative relative to the bands 4 and 5 with $\frac{t_p}{t_o}$=0.1,$\frac{\Delta_O}{t_o}$=-0.5 . The gray line indicates the position of the crossing between the band 4 and 5 increasing the value of $\lambda$ for $\frac{\Delta_I}{t_o}$=0.1. From left to right the parameters are: $\frac{\lambda}{t_o}$=0.7, $\frac{\lambda}{t_o}$=0.8, $\frac{\lambda}{t_o}$=0.86, and $\frac{\lambda}{t_o}$=0.95}
\label{tb4b5m05}
\end{figure}

At this point, we would like to point out that the evolution of the nodal loops is not always the same in the parameters space. If we refer to the phase diagrams \ref{pdb4b5m05gxxm}, for example, we can fix the value of  $\frac{\Delta_I}{t_o}$=0.1 and draw a "vertical" line in the phase diagrams; by increasing $\lambda$, it is evident that the value of $\mathscr{I}_{\Gamma X}$ goes from 0 to 1, while $\mathscr{I}_{\Gamma M}$ goes from 1 to 0 on the $\Gamma$M phase diagram for 0.7$\le \frac{\lambda}{t_o} \le$0.95 and the value of $\mathscr{I}_{XM}$ remains equal to 1. This corresponds to a transition of the kind 011 $\rightarrow$ 101. In this case, the evolution of the nodal loop is obtained by increasing the value of $\lambda$ and it is shown in Fig. \ref{tb4b5m05}. 
In the first panel on the left, $\frac{\lambda}{t_o}$=0.7 and the nodal line is initially constituted by a pocket around the M point, recovering the layer inversion parity sequence 011. Increasing the value of $\lambda$, this pocket moves along the diagonal direction, changing its concavity. In the second panel, $\frac{\lambda}{t_o}$=0.8, in the third panel $\frac{\lambda}{t_o}$=0.86 and the topology of the nodal loop has changed; in the last panel on the right, $\frac{\lambda}{t_o}$=0.95 and the pocket has formed around the $\Gamma$ point. There is not a zero along the XM direction, while a zero along the $\Gamma$X direction appears, so that we find a final configuration 101 for the inversion parity number.

\begin{figure}[t]
\noindent \begin{centering}
\includegraphics[height=2cm]{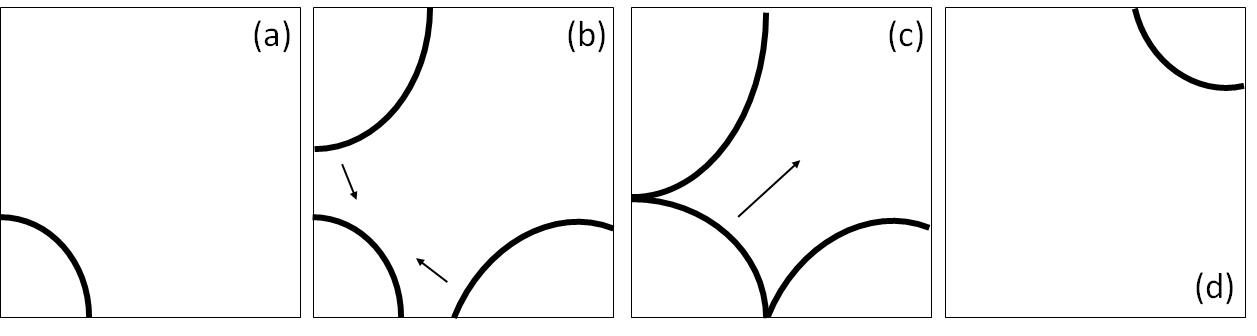}
\par\end{centering}
\caption{Schematic evolution of type A nodal loop transitions.}
\label{typeA}
\end{figure}

\begin{figure}[t]
\noindent \begin{centering}
\includegraphics[height=2cm]{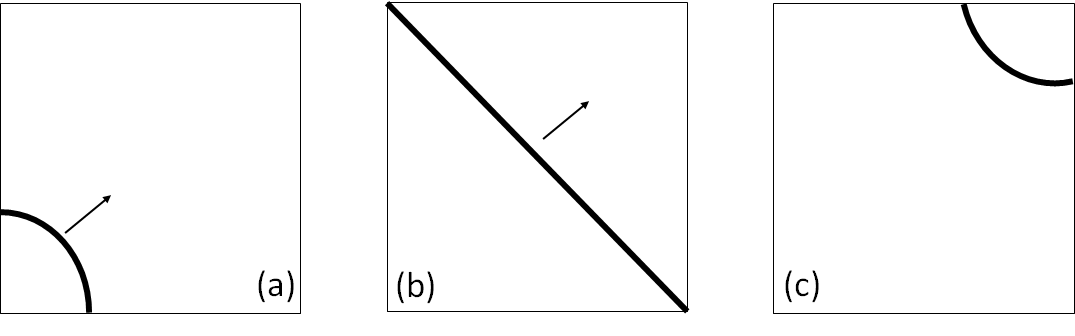}
\par\end{centering}
\caption{Schematic evolution of type B nodal loop transitions.}
\label{typeB}
\end{figure}

In summary, for the second type of transition the nodal line is not separated along a specific direction but it moves continuously in the BZ .\\
In light of the obtained results, we individuate several distinctive evolutions of the nodal loops which can be shortly described in the following list:

\begin{itemize}
\item type A: a nodal loop is located around a high symmetry point (panel (a) of Fig. \ref{typeA}). By varying  the value of one of the two parameters reported on the phase diagram, another pocket which is located around another high symmetry point emerges (panel (b) of Fig. \ref{typeA}) . These nodal loops get closer until they merge (panel (c) of Fig. \ref{typeA}). During this evolution, the value of the $\mathscr{I}$ is zero due to a double band inversion along the direction of the BZ that connects those points. The resulting nodal line subsequently moves far from that direction and approaches the remaining point (panel (d) of Fig. \ref{typeA}).
 
\item type B: we start from a pocket around a high symmetry point (panel (a) of Fig. \ref{typeB}). By varying the value of one of the two parameters the nodal line goes far from the original point and passes through the adjacent corners (panel (b) of Fig. \ref{typeB}); finally, it evolves around the point which is at the opposite site (panel (c) of Fig. \ref{typeB}). In doing that, it changes its concavity by reversing its sign.
\end{itemize}

The above investigated transitions can also occur for other electron filling configurations. The results are summarized in the Table \ref{alltrans}. The first column of the table includes all pairs of bands which manifests the transition, then we specify the parity value relative to the transition and the type of transition, according to the previous list; then, the following columns indicate the region of parameters for which the transition can be observed.

\begin{table}[!h]
\caption{Nodal loops transitions}
\begin{center}
\begin{tabular}{cccccc}
\hline
\rowcolor[gray]{.9} Bands & Crossing&Type&$\frac{\Delta_O}{t_0}$&$\frac{\lambda}{t_0}$& $\frac{\Delta_I}{t_o}$ \\[2mm]
\rowcolor[gray]{.8} b3b4 & 011$\rightarrow$101&type B&0.5&0.2&-0.48$\le\frac{\Delta_I}{t_o}\le$-0.2 \\[2mm]

\rowcolor[gray]{.9} b4b5 & 011$\rightarrow$101&type A&0.5&0.2& 0.34$\le\frac{\Delta_I}{t_o}\le$0.35 \\[2mm]

\rowcolor[gray]{.9} b4b5 & 011$\rightarrow$110&type A&0.5&0.8& 0.05$\le\frac{\Delta_I}{t_o}\le$0.1 \\[2mm]
\rowcolor[gray]{.8} b7b8 & 011$\rightarrow$101&type A&0.5& 0.65$\le\frac{\lambda}{t_o}\le$0.77&0.2 \\[2mm]

\rowcolor[gray]{.9} b4b5 & 011$\rightarrow$101&type B&-0.5& 0.7$\le\frac{\lambda}{t_o}\le$0.95&0.1 \\[2mm]


\rowcolor[gray]{.8}b5b6 & 011$\rightarrow$101&type B&-0.5& 0.1&-0.42$\le\frac{\Delta_I}{t_0}\le$-0.37 \\[2mm]


\rowcolor[gray]{.9} b6b7 & 011$\rightarrow$101&type B&-0.5&0.2&0.2$\le\frac{\Delta_I}{t_0}\le$0.49 \\[2mm]


\rowcolor[gray]{.8}b7b8 & 110$\rightarrow$101&type A&-0.5& 0.95$\le\frac{\lambda}{t_o}\le$1&0 \\[2mm]


\end{tabular}
\end{center}
\label{alltrans}
\end{table}

\section{Conclusions}

We have investigated the character of nodal line electronic structure in two-dimensional trilayered systems marked by local multi-orbital bands and spin-orbit coupling which are protected by the symmetry associated to the inversion layer parity. In the regime where the local energy splitting is more relevant than the 2D electron kinetic energy, we demonstrate that the spin-orbit coupling and the layer dependent crystal field potential can drive different types of {\it topological} transitions accompanied by a change of the nodal loops structure. Indeed, a single nodal loop winding around one of the high symmetry points can be converted into a pocket around another hhigh symmetry point of the Brillouin zone or it can splits in two loops. In order to fully characterize the nodal structure of the system we employ a criterion which is based on the evaluation of the band parity in the high symmetry positions of the Brillouin zone. Since the overall inversion parity cannot be changed, the nodal loops transitions are associated to a changeover of the crossings where the inversion layer parity is exchanged in two of the three symmetry positions. Finally, we also observe that the transitions in the nodal loops can be generally driven by tuning the crystal field potentials and that the relative sign between the inner and outerlayers does not play a crucial role (see Table I).

\acknowledgments
W.B. acknowledges support by the European Horizon 2020 research and
innovation programme under the Marie-Sklodowska-Curie grant agreement
No. 655515 
No. 2012/04/A/ST3/00331.
We are thankful to P. Barone and P. Gentile for valuable discussions and fruitful crticisms.

\end{document}